\documentclass[manuscript,acmsmall,screen]{acmart}
\renewcommand\footnotetextcopyrightpermission[1]{} 
\usepackage{algorithm}
\usepackage[]{algpseudocodex}

\usepackage{multirow}
\usepackage{xfrac}

\usepackage[binary-units]{siunitx}
\DeclareSIUnit\operations{OPS}
\DeclareSIUnit\instructions{I}
\DeclareSIUnit\byte{Byte}
\DeclareSIUnit\edges{edges}
\DeclareSIUnit\dependencies{dep.}
\DeclareSIUnit\node{node}

\usepackage{soul}
\newcommand{\meanvardeltatiteration}{\overline{\operatorname{Var}\left(\Delta {t_\text{iteration}}_{[k_\text{stop},k]}\right)}}
\newcommand{\meanvardeltatoverlap}{\overline{\operatorname{Var}\left(\Delta {t_\text{overlap}}_{[k_\text{stop},k]}\right)}}

\AtBeginDocument{%
  }

\copyrightyear{2024}

\acmConference[arXiv]{Make sure to enter the correct conference title from your rights confirmation emai}{September, 2024}{}
\acmPrice{15.00}
\acmISBN{978-1-4503-XXXX-X/18/06}

\setcopyright{none}
\begin{document}

\title{Automatic Generation of Fast and Accurate Performance Models for Deep Neural Network Accelerators}

\author{Konstantin Lübeck}
\orcid{0000-0002-2701-5881}
\affiliation{%
\institution{University of Tübingen}
  \streetaddress{Sand 13}
  \city{Tübingen}
  \country{Germany}
  \postcode{72076}
}

\author{Alexander Louis-Ferdinand Jung}
\orcid{0000-0001-5702-5768}
\affiliation{%
\institution{University of Tübingen}
  \streetaddress{Sand 13}
  \city{Tübingen}
  \country{Germany}
  \postcode{72076}
}

\author{Felix Wedlich}
\orcid{0009-0007-2632-7078}
\affiliation{%
\institution{University of Tübingen}
  \streetaddress{Sand 13}
  \city{Tübingen}
  \country{Germany}
  \postcode{72076}
}

\author{Mika Markus Müller}
\orcid{0009-0003-3471-7560}
\affiliation{%
\institution{University of Tübingen}
  \streetaddress{Sand 13}
  \city{Tübingen}
  \country{Germany}
  \postcode{72076}
}

\author{Federico Nicolás Peccia}
\orcid{0000-0002-3587-0415}
\affiliation{%
\institution{FZI Research Center for Information Technology}
  \streetaddress{Haid-und-Neu-Str. 10–14}
  \city{Karlsruhe}
  \country{Germany}
  \postcode{76131}
}

\author{Felix Thömmes}
\orcid{0009-0007-0885-5424}
\affiliation{%
\institution{FZI Research Center for Information Technology}
  \streetaddress{Haid-und-Neu-Str. 10–14}
  \city{Karlsruhe}
  \country{Germany}
  \postcode{76131}
}

\author{Jannik Steinmetz}
\orcid{0009-0003-7193-5620}
\affiliation{%
\institution{University of Tübingen}
  \streetaddress{Sand 13}
  \city{Tübingen}
  \country{Germany}
  \postcode{72076}
}

\author{Valentin Biermaier}
\orcid{0009-0007-9428-9085}
\affiliation{%
\institution{University of Tübingen}
  \streetaddress{Sand 13}
  \city{Tübingen}
  \country{Germany}
  \postcode{72076}
}

\author{Adrian Frischknecht}
\orcid{0000-0002-1795-0948}
\affiliation{%
\institution{University of Tübingen}
  \streetaddress{Sand 13}
  \city{Tübingen}
  \country{Germany}
  \postcode{72076}
}

\author{Paul Palomero Bernardo}
\orcid{0000-0002-6642-3976}
\affiliation{%
\institution{University of Tübingen}
  \streetaddress{Sand 13}
  \city{Tübingen}
  \country{Germany}
  \postcode{72076}
}

\author{Oliver Bringmann}
\orcid{0000-0002-1615-507X}
\affiliation{%
\institution{University of Tübingen}
  \streetaddress{Sand 13}
  \city{Tübingen}
  \country{Germany}
  \postcode{72076}
}

\renewcommand{\shortauthors}{Lübeck et al.}

\begin{abstract}
Implementing Deep Neural Networks (DNNs) on resource-constrained edge devices is a challenging task that requires tailored hardware accelerator architectures and a clear understanding of their performance characteristics when executing the intended AI workload. To facilitate this, we present an automated generation approach for fast performance models to accurately estimate the latency of a DNN mapped onto systematically modeled and concisely described accelerator architectures. 

Using our accelerator architecture description method, we modeled representative DNN accelerators such as Gemmini, UltraTrail, Plasticine-derived, and a parameterizable systolic array. Together with DNN mappings for those modeled architectures, we perform a combined DNN/hardware dependency graph analysis, which enables us, in the best case, to evaluate only 154 loop kernel iterations to estimate the performance for 4.19 billion instructions achieving a significant speedup. We outperform regression and analytical models in terms of mean absolute percentage error (MAPE) compared to simulation results, while being several magnitudes faster than an RTL simulation.
\end{abstract}




\maketitle

\section{Introduction}
Edge devices provide lower bandwidth demands, lower response times, higher autonomy, better privacy, less cost, and higher energy efficiency than data centers when processing data-intensive tasks such as Deep Neural Networks. Therefore, more and more hardware vendors enter the market for edge devices tailored explicitly for the execution of DNNs, offering a large variety of parameterizable accelerator architectures. However, selecting the most suitable accelerator architecture for a given application is arduous since reliable performance metrics can often only be gathered using time-consuming simulators or data sheets that provide peak \si{\operations\per\second} for only a few selected DNNs so that these numbers are hardly transferrable to other DNN architectures, using spreadsheet calculations, due to their vast hyperparameter space.

Additionally, during the early design phase of an accelerator, it is crucial to compare different variants with each other quickly. On the hardware side, these variants range from changes in the number of available multiply-accumulate units to entirely different data paths. On the software side, hardware-aware Network Architecture Search (NAS) can explore different hyperparameters and data reuse strategies for mapping DNNs onto an accelerator. Building accurate performance models for all those variants and combinations is only feasible with a fast and automated approach.

Therefore, we propose an automatic performance estimation approach enabling fast yet accurate DNN accelerator modeling and evaluation on different abstraction levels for hardware and software.

This paper makes the following key contributions:
\begin{itemize}
    \item  An introduction to the Abstract Computer Architecture Description Language (ACADL), which allows us to model and evaluate a broad range of accelerator architectures with various architectural parameters at different abstraction levels.
	\item We automatically generate an Architectural Instruction Dependency Graph (AIDG) given an accelerator architecture described in ACADL and a DNN mapping. This is a novel approach where both hardware and software are represented in a flexible and abstract manner, while other performance estimation methods mainly focus on software (e.g., instruction set simulators) or hardware (e.g., register-transfer level (RTL) simulators). 
	\item We propose a fast evaluation methodology of AIDGs to accurately estimate the performance of DNNs mapped onto accelerator architectures, where only as few as \num{0.0001}\% of all loop kernel iterations have to be analyzed to get an accurate performance estimation, matching RTL simulation accuracy and outperforming literature-reported regression and analytical models.
\end{itemize}

\section{Related Work}
Several studies have been carried out to estimate the performance of DNNs executed on accelerator architectures. Those studies can be divided into analytical and machine learning models with varying accuracy and estimation speeds.

Bouzidi et al. \cite{bouzidi2021} compare five different machine learning models for performance estimation of DNNs mapped onto edge GPUs. They executed different DNN architectures with varying hyperparameters, measured their end-to-end latency and gathered at least \num{200000} samples per platform to train different estimators. The MAPE for the best regression models ranges between 14.73\% and 7.67\%.

In PreVIous \cite{previous2020}, a specially designed DNN composed of different layers with varying parameters is executed on the CPUs of single-board computers. During the execution of this DNN, the end-to-end latency and energy consumption for each layer are measured. The measurements are used together with the layer parameters to train linear regression models for execution time and energy consumption estimation. In terms of execution time, PreVIous achieves a MAPE between 3.25\% and 7.92\%. However, collecting the data is time-consuming and requires the hardware for which the estimation should be made.

ANNETTE \cite{annette2021} uses micro-kernel benchmarks and multi-layer benchmarks of DNNs executed on an FPGA and an ASIC accelerator to characterize the performance of those platforms. Similar to PreVIous special DNNs are used to measure the execution time of several different layers in one benchmark run. The layer execution time measurements are combined with refined roofline models, incorporating utilization efficiency, for each layer which are used when no measurements for a layer exist to estimate the execution time of a whole DNN executed on one of the selected accelerators reaching a MAPE between 3.47\% and 7.44\%.

Timeloop \cite{timeloop2019} proposes an analytical modeling framework for execution time and energy consumption based on coarse textual descriptions of accelerator architectures and the mapping of convolutional and fully-connected layers. Compared to three real-world accelerators, an average accuracy in terms of execution time of 95\% is achieved. Timeloop does not consider pipeline stalls, resource conflicts and instruction-level parallelism, which can lead to an accuracy as low as 78\%.

Aladdin \cite{aladdin2014} and its extension \cite{wang2022} construct an unconstrained data dependency graph for an algorithm. Then hardware properties of an accelerator architecture, the algorithm should be mapped onto, constrain this data dependency graph resulting in an average execution time estimation error of 0.9\% for a simple but parameterizable accelerator (Aladdin), and an estimation error of 7.1\% for Eyeriss, and 5.6\% for the DSIP accelerator (extension).

Our estimation approach provides a systematic way to model accelerator architectures that cover various parameterizable design alternatives on different hardware and software abstraction levels, ranging from scalar operations up to fused tensor operations. We show the generality and accuracy of our proposed approach by modeling four architecturally different accelerators and estimating the performance of those using three state-of-the art DNNs tailored for edge devices and comparing it with literature-reported regression models \cite{bouzidi2021}, refined roofline models \cite{annette2021}, and Timeloop \cite{timeloop2019}.

\section{Proposed Approach}
\begin{figure}[htbp]
	\centerline{\includegraphics[width=1.0\linewidth]{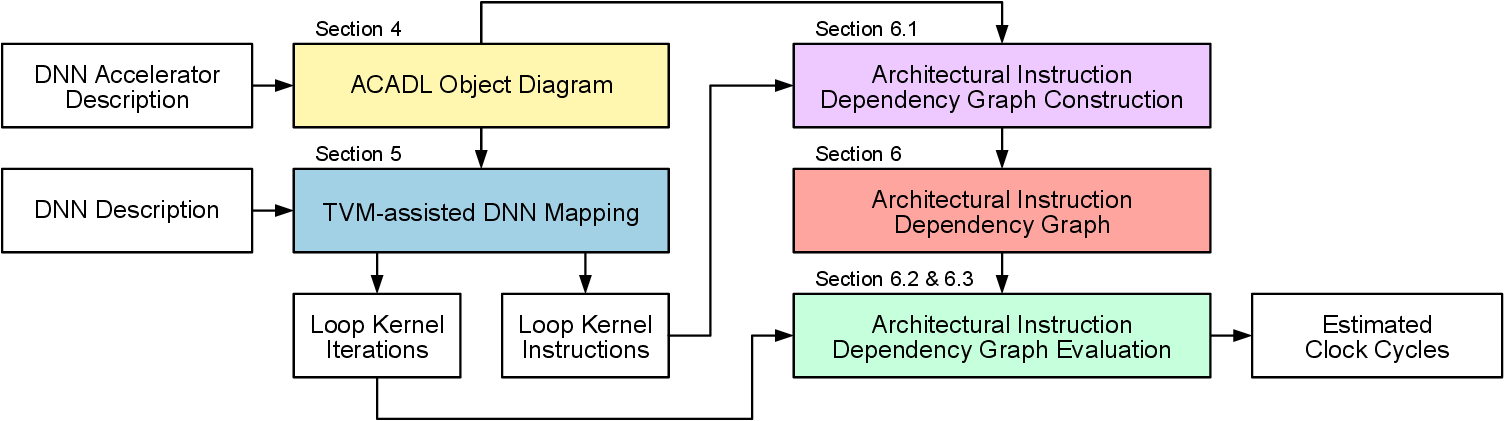}}
	\caption{Overview of the automatic DNN accelerator performance model generation approach.}
	\label{fig:overview}
\end{figure}

Fig.~\ref{fig:overview} shows our proposed automatic DNN accelerator performance model generation approach. The foundation of our performance model generation is the \textbf{Abstract Computer Architecture Description Language (ACADL)} introduced in section \ref{sec:acadl}. ACADL allows us to model various parameterizable accelerator architectures and capture how data and instructions propagate through them on different abstraction levels.

A \textbf{TVM-assisted Deep Neural Network mapping} presented in section \ref{sec:dnn_mapping} for a given DNN accelerator architecture described in ACADL generates loop kernel instructions together with the number of iterations for each DNN layer.

Each loop kernel instruction is propagated through the ACADL object diagram to construct an Architectural Instruction Dependency Graph (AIDG) in the \textbf{Architectural Instruction Dependency Graph Construction} phase described in section \ref{sec:aidg_construction}. This directed acyclic graph captures the structural and data dependencies between instructions occupying hardware modules, which represents the performance model for a DNN/hardware mapping.

The \textbf{Architectural Instruction Dependency Graph Evaluation} in section \ref{sec:aidg_evaluation} uses the generated AIDG and the calculated loop kernel iterations. Depending on the loop kernel instructions, the loop must be executed only a few times to estimate the end-to-end latency of a whole DNN layer. Since in consecutively executed iterations, only the memory addresses change, it is only necessary to analyze the first few iterations until a stable end-to-end latency of a single iteration is established. This stable end-to-end latency is then multiplied by the amount of loop kernel iterations minus the loop kernels that have already been analyzed, resulting in the end-to-end latency in clock cycles of a whole DNN layer. The sum of all DNN layer latencies is the end-to-end latency for a whole DNN.

As a running example for our proposed approach, we use a parameterizable systolic array accelerator architecture in each section, from modeling (section~\ref{sec:acadl_modeling_examples}), over DNN mapping (section~\ref{sec:dnn_mapping}) to latency estimation (section~\ref{sec:aidg_evaluation}), and results (section~\ref{sec:systolic_array_results}).

\section{Abstract Computer Architecture Description Language}
\label{sec:acadl}
\noindent Computer architectures are almost exclusively communicated using block diagrams. Each block in a block diagram describes the function of the computer architecture module it represents, while arrows are used to depict how data is exchanged between different modules. However, modeling computer architectures on the system level is mainly done using hardware description languages like VHDL and Verilog or electronic system-level languages such as SystemC. 

Especially in the early design phase of new computer architectures, it is crucial to evaluate the performance of different design alternatives. However, using the aforementioned hardware description languages requires a lot of expert knowledge and time to produce working design alternatives, leading to low productivity. Therefore, we use the Abstract Computer Architecture Description Language (ACADL) \cite{acadl2024} that does not require extensive hardware design knowledge. 

\begin{figure}[htbp]
	\centerline{\includegraphics[width=1.0\linewidth]{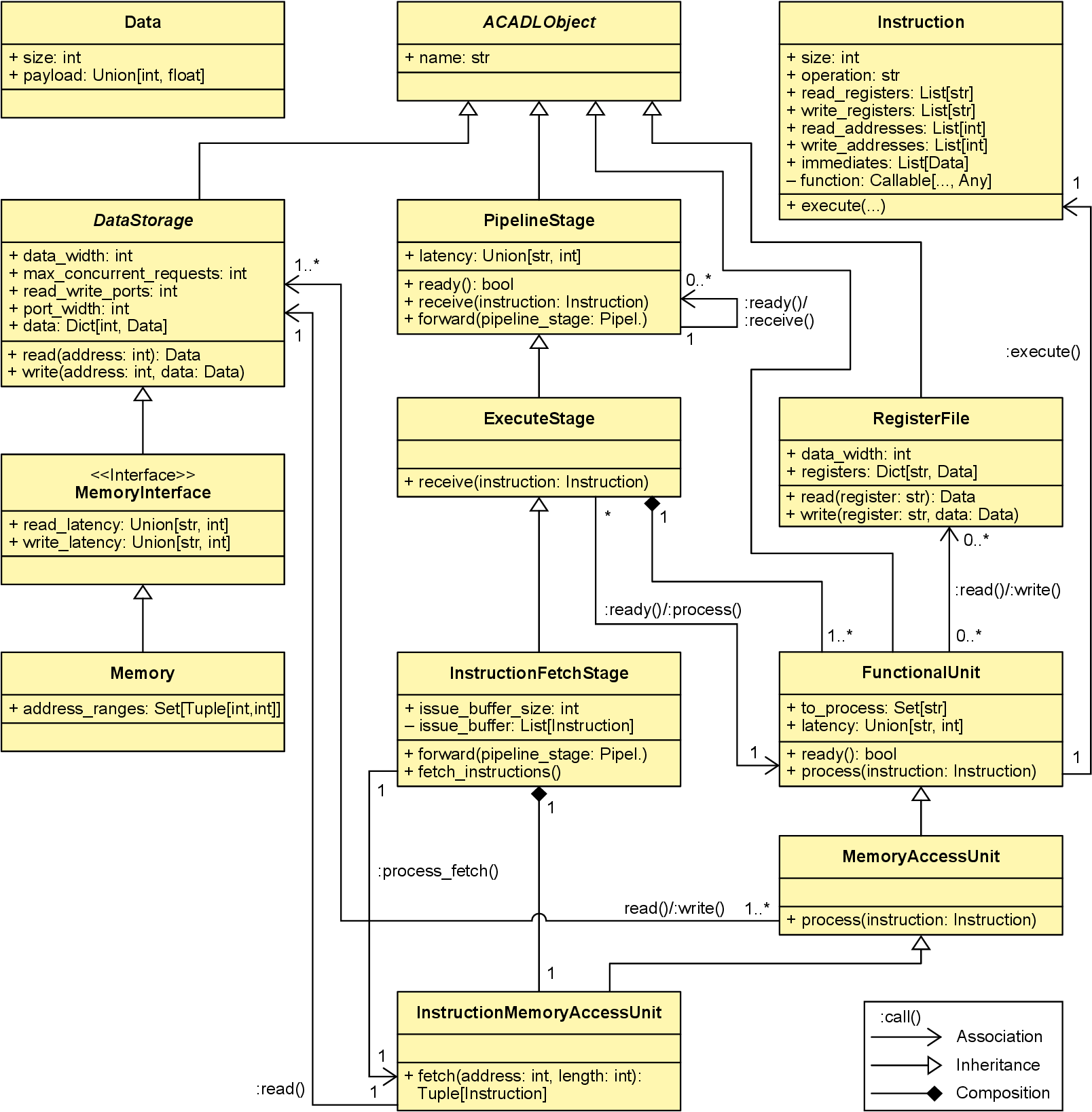}}
    \caption{Abstract Computer Architecture Description Language class diagram.}
	\label{fig:acadl_class_diagram}
\end{figure}

ACADL is an object-oriented language that defines twelve classes and one interface that describe the basic building blocks of computer architectures. The dependencies and relations of those classes are depicted in Fig.~\ref{fig:acadl_class_diagram} using the Unified Modeling Language (UML) \cite{uml2017} notation standard for class diagrams. Computer architectures modeled with ACADL are instruction-centric, meaning that any architectural state change is triggered by an instruction. All instructions originate in an instruction memory and are forwarded to a functional unit supporting the instruction's operation and register and memory accesses. The time in clock cycles from loading an instruction from the instruction memory until it reaches a functional unit that carries out its operation is accumulated. This accumulated time is the end-to-end latency of a single instruction. Several instructions can be loaded, issued, forwarded, and executed simultaneously, allowing for modeling multiple issue and parallel computer architectures with ACADL.

\subsection{Types and Classes}\label{acadl_types_and_classes}
The ACADL class diagram in Fig.~\ref{fig:acadl_class_diagram} shows the dependencies between all classes. An association describes that a class calls the method of another class. Inheritance describes that a derived class inherits all attributes and methods from its base class. A derived class can add its own unique attributes and methods. In Fig.~\ref{fig:acadl_class_diagram} we only list the attributes and methods in the base class and only list the added attributes in methods in the derived class. The composition expresses that a class contains instances of other classes as attributes.

\textbf{bool, int, str, List, Set, Union, Tuple, Any, Callable, ...} are the data types of ACADL based on the \mbox{Python 3.10} typing system \cite{python3typing2022}.

\textbf{Association} describes the relation between two or more instantiated objects of classes that allows one object (caller) to cause another one to perform an action (callee), while \textbf{:call()} signifies the callee's function that is called by the caller.

\textbf{Inheritance} allows a class to base its implementation and attributes upon another base class.

\textbf{Composition} defines a relationship between instantiated objects of classes that implies that one or more objects are part of a single composite object.

\textbf{latency} describes a time delta in clock cycles. It can be specified as an integer value or a string containing a function that is evaluated during the performance estimation.

\textbf{ACADLObject} is the virtual base class for every computer architecture module modeled in ACADL. It only has the attribute \texttt{name}, the unique identifier for each object.

\textbf{Data} represents any numerical data stored in memories, registers, and immediate values of instructions. \texttt{size} is the data size in bits. \texttt{payload} is the data itself, which is only used for an optional functional simulation.

\textbf{Instruction} has several attributes describing which registers (\texttt{read\_registers}, \texttt{write\_registers}) and memory addresses (\texttt{read\_addresses}, \texttt{write\_addresses}) and immediate values (\texttt{immediates}) are accessed when it is executed. \texttt{operation} is the instruction's mnemonic, while \texttt{function} contains how data is manipulated. \texttt{execute()} calls \texttt{function} when the instruction is processed by a FunctionalUnit. Furthermore, an instruction is not limited to fine-grained operations such as addition or multiplication. An instruction can also carry out complex operations like matrix-matrix multiplication or Fourier transformations. This enables modeling at different abstraction levels.

\textbf{PipelineStage} is responsible for forwarding instructions inside a computer architecture. \texttt{receive()} is called by \texttt{forward()} of another PipelineStage which forwards an instruction between two PipelineStages. An Instruction can only be forwarded if the receiving PipelineStage is \texttt{ready()}. The amount of clock cycles an Instruction resides inside a PipelineStage before it is forwarded is indicated by \texttt{latency}.

\textbf{RegisterFile} contains \texttt{registers} that maps the unique register name to a value. \texttt{data\_width} is the size of each register in bits, while \texttt{read()} and \texttt{write()} provide access to the stored values.

\textbf{FunctionalUnit} executes an Instruction that is passed to \texttt{process()} and changes the architectural state by changing register contents using \texttt{read()} and \texttt{write()} of RegisterFiles. A RegisterFile can be read-only, write-only, or readable and writable depending on the \texttt{:read()}/\texttt{:write()} associations connecting FunctionalUnits and RegisterFiles. FunctionalUnits can only process Instructions whose \texttt{operation} is in \texttt{to\_process} and if the FunctionalUnit has read/write access to the \texttt{read\_registers} and \texttt{write\_registers}. Processing a supported Instruction takes \texttt{latency} clock cycles after all data dependencies from previous Instructions are resolved.

\textbf{ExecuteStage} inherits from PipelineStage and overrides \texttt{receive()} and additionally contains FunctionalUnits. When an ExecuteStage receives an Instruction, it checks all its contained FunctionalUnits if one of them supports processing the Instruction. The check involves if \texttt{operation} is in \texttt{to\_process} and if the \texttt{read\_registers} and \texttt{write\_registers} are accessible by the FunctionalUnit. When a contained FunctionalUnit supports the Instruction, the ExecuteStage passes the Instruction to \texttt{process()} and the ExecuteStage's \texttt{latency} is not accumulated.

\textbf{DataStorage} is the virtual base class for all memories. \texttt{data\_width} describes the bit-length of one data word. \texttt{max\_concurrent\_requests} is the maximum amount of read and write requests that can be handled at the same time while \texttt{read\_write\_ports} describes how many MemoryAccessUnits can be connected to the DataStorage. \texttt{port\_width} is the amount of data words that can be accessed during a single memory transaction. A $\texttt{port\_width} > 1$ allows for reading or writing several data words at once. \texttt{data} maps a memory address to a data word.

\textbf{MemoryInterface} adds \texttt{read\_latency} and \texttt{write\_latency} to DataStorages which describe how many clock cycles a read or write transaction takes when \texttt{read()} or \texttt{write()} are called. 

\textbf{Memory} stores data at addresses included in \texttt{address\_ranges}. Together with RegisterFiles, Memories enable modeling of complex memory hierachies with different latencies for each level.

\textbf{MemoryAccessUnit} inherits from FunctionalUnit and overrides \texttt{process()} to access Memories and RegisterFiles. 

\textbf{InstructionMemoryAccessUnit} inherits from MemoryAccessUnit and adds \texttt{fetch()}, which reads \texttt{length} Instructions starting at \texttt{address} from an instruction Memory.

\textbf{InstructionFetchStage} inherits from ExecuteStage and overrides \texttt{forward()}, which forwards an instruction residing in the \texttt{issue\_buffer}. It is possible to forward multiple Instructions in the same clock cycle. \texttt{issue\_buffer\_size} describes how many Instructions fit into the \texttt{issue\_buffer} and thereby marks the maximum number of Instructions that can be issued in a single clock cycle. \texttt{fetch\_instructions()} is called in each clock cycle as long as the \texttt{issue\_buffer} is not full.

The performance of a computer architecture is dependend on the latencies and interactions of different hardware modules. ACADL provides classes that represent the basic building blocks of a computer architecture together with their dependencies and a precise latency semantic. The virtual base class ACADLObject for all hardware modules does not have a latency attribute, because not all classes derived from ACADLObject can be described with a single latency. For example, all classes derived from MemoryInterface have a \texttt{read\_latency} and a \texttt{write\_latency} as those two latencies can be very different. Additionally, the RegisterFile class does not have a latency attribute because all register access latencies are implicitly described by the latency of FunctionalUnits that write and read to and from RegisterFiles. This is a deliberate decision because modeling a computer architecture in ACADL is instruction-centric and the latency of processing an instruction accessing registers is described through the latency of a FunctionalUnit. Lastly, adding a latency attribute to the virtual base class ACADLObject would limit the expendability of ACADL because all derived classes must inherit the latency attribute, even though they might not require one.

\subsection{General Modeling Workflow}
Creating an ACADL object diagram for an accelerator architecture is the foundation of our proposed performance evaluation approach. In this section, we introduce two methods for obtaining an ACADL object diagram. 

The first method is the top-down modeling of an accelerator. This modeling method is to be used when exploring the design of a new architecture. Using the envisioned accelerator capabilities supported operations are formulated. Together with structural information of the accelerator hardware a high-level ACADL object diagram can be created. Mapping DNNs onto the ACADL object diagram together with the proposed performance evaluation allows to validate the high-level design. This validated ACADL object diagram can than be decomposed into smaller modules and validated again. By iterative refinement and validation, a detailed ACADL object diagram for an accelerator is obtained.

The second method is modeling accelerator architectures bottom-up which is to be used for obtaining an ACADL object diagram for an existing accelerator. This method requires some knowledge of the hardware that implements the accelerator. An input for the bottom-up modeling is an architectural block diagram together with timing information on different operations running on the accelerator. The timing information can be obtained from the accelerator datasheet or through microbenchmarks. The architectural block diagram can often be translated into an ACADL object diagram without major adaptations. Section \ref{sec:acadl_modeling_examples} provides two detailed examples for obtaining an ACADL object diagram from architectural block diagrams on two abstraction levels. Additionally, in section \ref{sec:gemmini_results} the ACADL object diagram for the Gemmini accelerator is presented which was also obtained from the architectural block diagram of the Gemmini accelerator.

The UML-based representation of the ACADL class diagram enables developers to use a graphical UML editor to create accelerator architecture object diagrams. A suitable export plugin would be required to automatically generate an analyzable ACADL model from such an object diagram.

Additionally, \cite{acadl2024} provides a detailed description on how to use ACADL to model AI hardware accelerators.

\subsection{Modeling Examples}
\label{sec:acadl_modeling_examples}

In this section, we present two different example accelerator architectures, to show how these can be modeled as an ACADL object diagram using their architectural block diagrams. While the first architecture is modeled in ACADL using scalar instructions, the second architecture is modeled using fused tensor instructions and thus addressing different abstraction levels. 

\begin{figure}[htbp]
	\centerline{\includegraphics[width=1.0\linewidth]{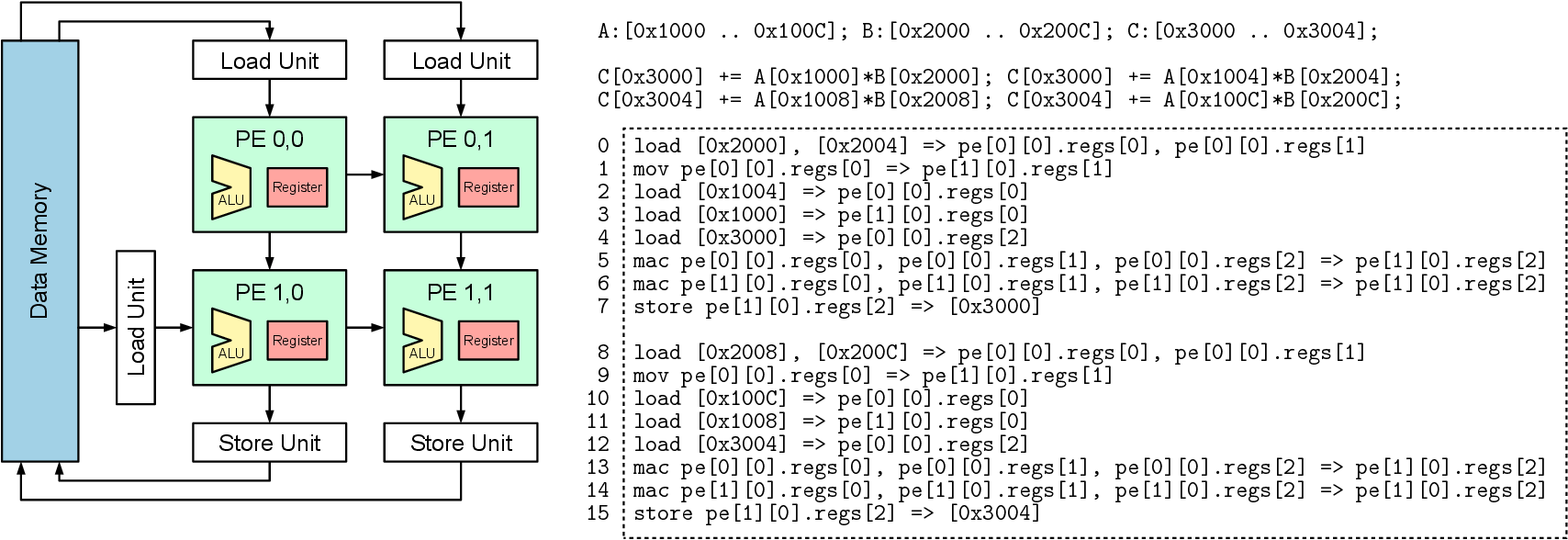}}
	\caption{Block diagram and example instructions for a for a 2$\times$2 systolic array.}
	\label{fig:systolic_array_block_diagram}
\end{figure}

The block diagram in Fig.~\ref{fig:systolic_array_block_diagram} depicts a 2$\times$2 systolic array. Each processing element (PE) consists of an ALU and an internal register file. PEs are able to move data from their internal register file into the register file of adjacent PEs at the bottom and the right-hand side. The PEs at the top row and leftmost column are connected to load units that read data from the data memory. The PEs in the bottom row are connected to store units that are able to write data back into the data memory. The instructions presented on the right-hand side of Fig.~\ref{fig:systolic_array_block_diagram} show two consecutive iterations of an element-wise multiplication of vectors A and B accumulated into vector C. 

\begin{figure}[htbp]
	\centerline{\includegraphics[width=\linewidth]{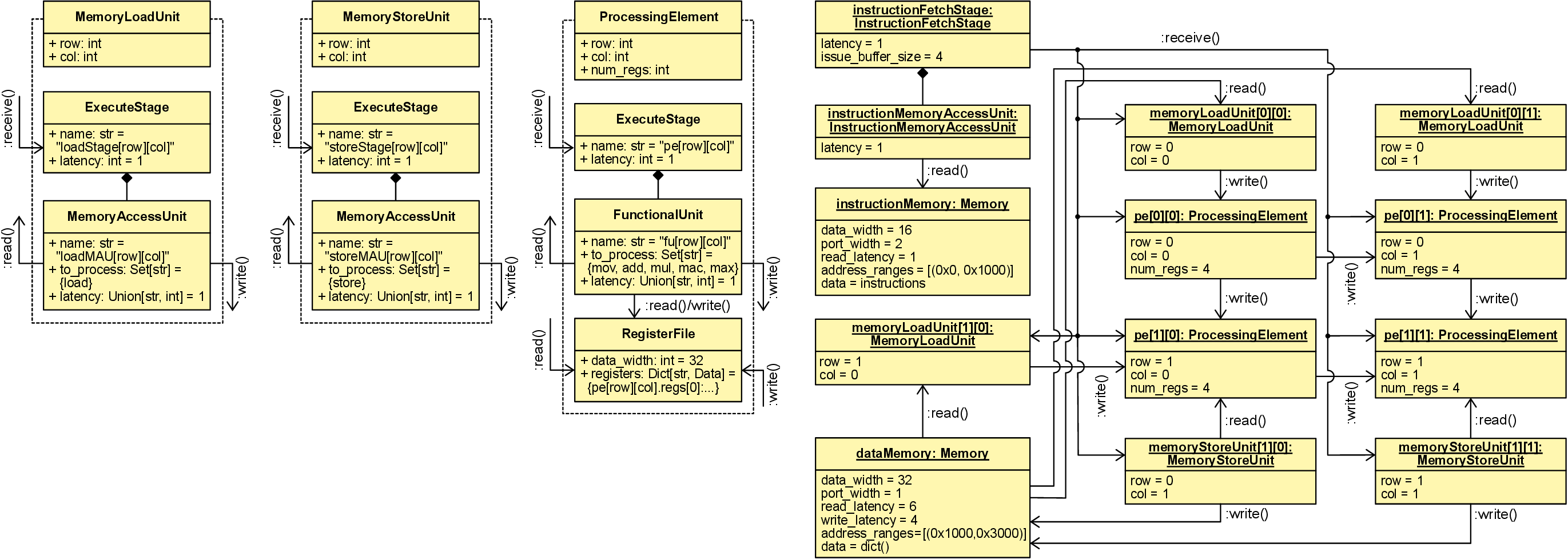}}
	\caption{Classes and ACADL object diagram for a 2$\times$2 systolic array.}
	\label{fig:systolic_array_acadl_object_diagram}
\end{figure}

Fig.~\ref{fig:systolic_array_acadl_object_diagram} shows additional classes and the ACADL object diagram for 2$\times$2 systolic array. The block diagram's load units, store units, and PEs are represented by the classes MemoryLoadUnit, MemoryStoreUnit, and ProcessingElement. Introducing those three classes, composed of ACADL classes, helps encapsulate their behavior and allows for reuse in the ACADL object diagram. Each of those classes contains an ExecuteStage and a FunctionalUnit, which are responsible for executing Instructions, while the ProcessingElement class also contains a RegisterFile. The data memory is represented by the Memory object \texttt{dataMemory} while the instantiated MemoryLoadUnits and MemoryStoreUnits (\texttt{memoryLoadUnit[0][0]}, \texttt{memoryLoadUnit[0][1]}, \texttt{memoryLoadUnit[1][0]}, \texttt{memoryStoreUnit[1][0]}, \texttt{memoryStoreUnit[1][1]}) manipulate the data memory through read and write associations. In the ACADL object diagram, the \texttt{instructionMemory}, which is a Memory object, is instantiated in which the instructions shown in Fig.~\ref{fig:systolic_array_block_diagram} reside. Using the \texttt{instructionMemory\-AccessUnit} and \texttt{instructionFetchStage}, instructions are read from the \texttt{instructionMemory} and forwarded to all FunctionalUnits.

Except for the \texttt{instructionMemoryAccessUnit}, \texttt{instructionFetchStage}, and \texttt{instructionMemory}, all objects in the ACADL object diagram can be found in the block diagram in Fig.~\ref{fig:systolic_array_block_diagram}. This shows that a given block diagram can be easily transferred into an ACADL object diagram.

\begin{figure}[htbp]
	\centerline{\includegraphics[width=1.0\linewidth]{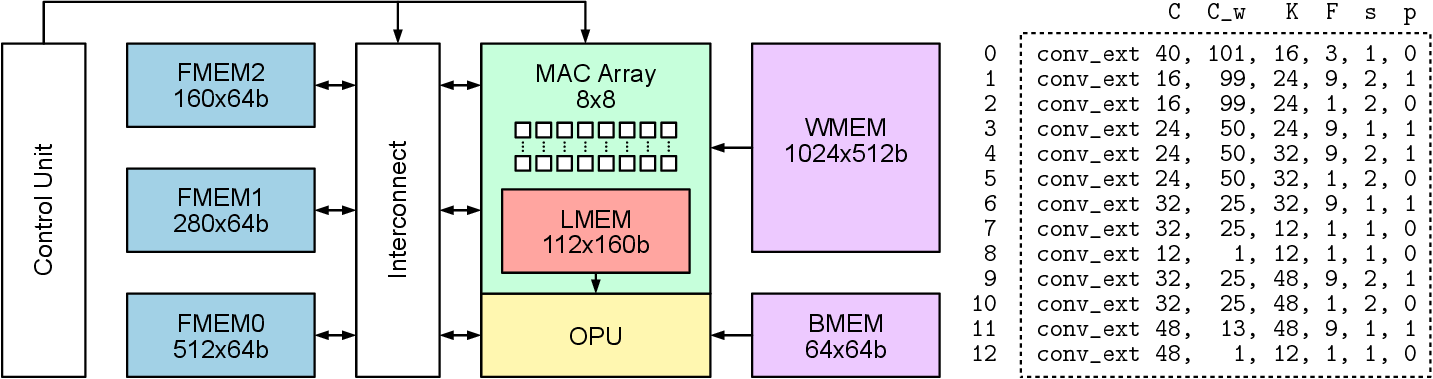}}
	\caption{Block diagram and example instructions for the UltraTrail accelerator \cite{ultratrail2020}.}
	\label{fig:ultratrail_block_diagram}
\end{figure}

The second example architecture is UltraTrail \cite{ultratrail2020}, an ultra-low power accelerator for 1-dimensional data processing developed by the University of Tübingen. The UltraTrail accelerator block diagram is depicted in Fig.~\ref{fig:ultratrail_block_diagram}. The accelerator is composed out of an $8{\times}8$ combinational MAC array computing convolutional layers. The MAC array is connected to four different memories supplying it with data. FMEM0-2 are feature memories storing the input features and the computed output features of the layers. The WMEM contains the filter weights for convolutional kernels. Because large convolutional layers do not fit entirely onto the MAC array, their computation is split. At the same time, partial results are stored in the local memory of the MAC array called LMEM. Due to the combinational layout of the MAC array, 64 MACs are executed in each clock cycle. After processing a complete convolutional layer, the output results are piped through the combinational output processing unit OPU that is responsible for adding a bias coming from BMEM, applying a ReLU activation, and average pooling. The authors describe a fused convolutional layer with an activation layer and/or pooling layer as CONV-EXT and provide an analytical performance model for the end-to-end latency for a CONV-EXT layer executed on the UltraTrail architecture. The parameters of the CONV-EXT layer from left to right are: number of input channels $C$, width of an input channel $C_w$, number of output channels $K$, filter width $F$, stride $s$, and padding enabled $p$ \cite{ultratrail2020}.

\begin{figure}[htbp]
	\centerline{\includegraphics[width=1.0\linewidth]{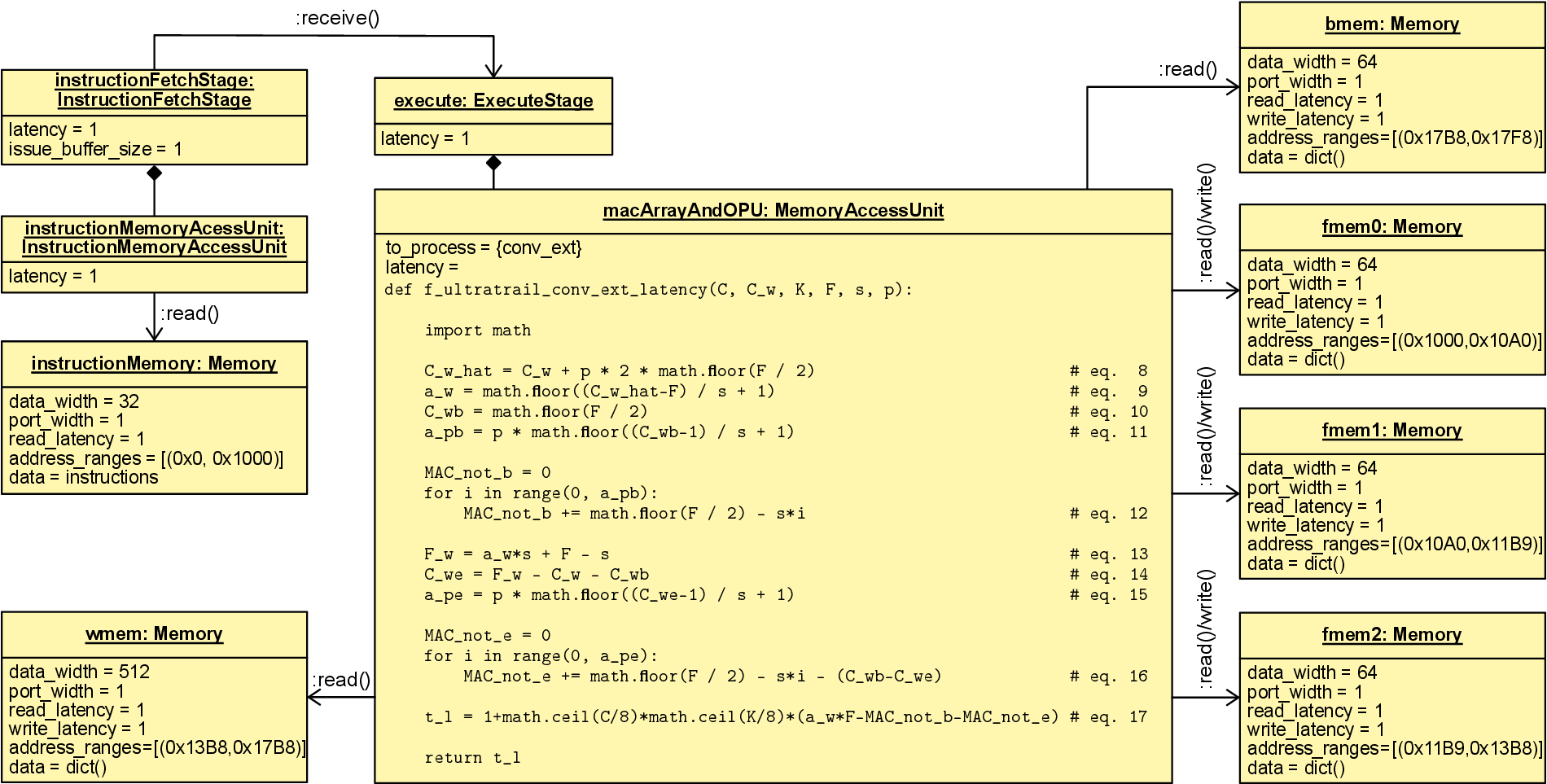}}
	\caption{ACADL object diagram for the UltraTrail accelerator.}
	\label{fig:ultratrail_acadl_object_diagram}
\end{figure}

Based on the UltraTrail CONV-EXT performance model we created an ACADL object diagram shown in Fig.~\ref{fig:ultratrail_acadl_object_diagram}. In contrast to the object diagram of the 2$\times$2 systolic array in Fig.~\ref{fig:systolic_array_acadl_object_diagram} where individual functional units are represented by different ACADL objects the MAC array together with the OPU are modeled as a single FunctionalUnit whose latency is set to the analytical performance model presented in \cite{ultratrail2020}. The input parameters for the analytical performance model are encoded in the \texttt{immediates} of the \texttt{conv\_ext} instructions shown in Fig.~\ref{fig:ultratrail_block_diagram}. Each time a \texttt{conv\_ext} instruction enters the \texttt{macArrayAndOPU} object the latency is evaluated using the parameters coming from the instruction's immediate values. Using the parameters provided in \cite{ultratrail2020} for the TC-ResNet8 \cite{tcresnet2019} we were able to match the end-to-end latency of the UltraTrail accelerator executing the TC-ResNet8 almost exactly to 22484 clock cycles and only overestimating it by three clock cycles which stem from the instruction fetch part that is not considered in the original performance model.

These two examples show how ACADL can be employed to model edge AI accelerator architectures at different abstraction levels, from fine-grained scalar modeling of individual functional units to coarse tensor models that bundle several hardware elements together. Nevertheless, ACADL is not limited to accelerator architectures and can also model processor pipelines and memory-mapped communication between CPUs and accelerators; however, this is beyond the scope of this paper. ACADL provides a high degree of flexibility in the software representation through instructions ranging from atomic operations, such as multiplication and addition, up to fused vector/matrix operations and loading or storing multiple data words in a single transaction, thereby enabling to model any DNN layer that can be expressed using those operations. These abstraction levels allow us to model complex accelerator architectures with a precise latency semantic and facilitate a shallow learning curve where a developer can start on a high abstraction level, such as the presented UltraTrail model, and can then iteratively refine the model down to single registers and multiply-accumulate units.

\subsection{Extendability and Limitations}
\begin{figure}[htbp]
	\centerline{\includegraphics[width=0.7\linewidth]{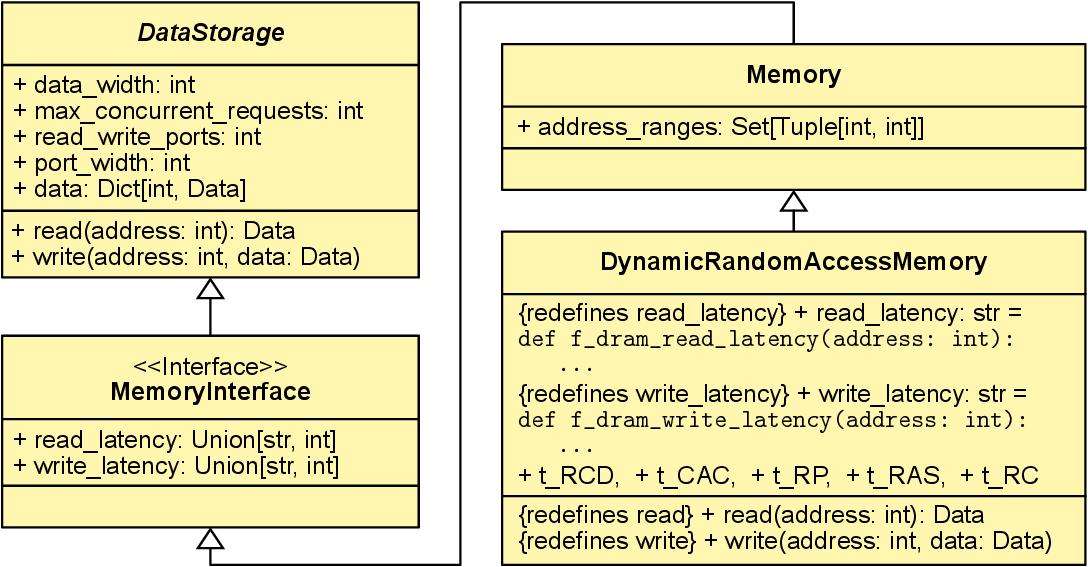}}
	\caption{Extension of the ACADL class diagram with DRAM.}
	\label{fig:acadl_extension}
\end{figure}

The ACADL class diagram presented in Fig.~\ref{fig:acadl_class_diagram} is easily extendable by adding new classes and inheriting from existing ones. Fig.~\ref{fig:acadl_extension} depicts how a new class for a dynamic random access memory (DRAM) model that inherits from Memory can be added to the ACADL class diagram. the Dynamic\-Random\-Access\-Memory has additional attributes for different access latencies for DRAM (\texttt{t\_RCD}, \texttt{t\_CAC}, \texttt{t\_RP}, and \texttt{t\_RAS}, \texttt{t\_RC}) and redefines \texttt{read\_latency} and \texttt{write\_latency} with stateful latency functions. The object-oriented paradigm facilitates the extendability of ACADL. Additionally, ACADL models could be generated from other languages such as Bluespec \cite{bluespec2009}, Chisel \cite{chisel2012}, or CoreDSL \cite{scale4edge2022}.

The version of ACADL presented in this paper does not support describing cache hierarchies, DRAM, and on-chip busses without extending the class diagram. We choose not to include those because many edge AI accelerator architectures heavily rely on SRAM and standard cell memory for their on-chip memory \cite{ultratrail2020, plasticine2017, eyerissv12017, yodann2016, dianao2014}. Additionally, there is currently no power semantic to describe the energy demand of a modeled computer architecture.

\section{Deep Neural Network Mapping}
\label{sec:dnn_mapping}

In the previous section, we have seen how we can model accelerator architectures on different abstraction levels using ACADL. To estimate the performance of an accelerator architecture, we also need to represent the DNN layers to be executed.

Generally speaking, all accelerator architectures exploit some kind of data reuse during the computation of DNN layers to reduce memory accesses, as those are typically the most time-consuming operations. For example, the UltraTrail architecture, presented in Fig.~\ref{fig:ultratrail_block_diagram}, unrolls the output channels $K$ in one dimension and the input channels $C$ in the other \cite{ultratrail2020}. A similar reuse pattern is used for the systolic array shown in Fig.~\ref{fig:systolic_array_block_diagram}.

The relevant difference is that the UltraTrail accelerator operates with tensor-level \texttt{conv\_ext} instructions, while the systolic array executes scalar instructions like \texttt{load} and \texttt{mac}. Therefore, the granularity of the representation of the DNN layer needs to match the abstraction level of the ACADL model.

In the case of the aforementioned systolic array, we use TVM's \cite{tvm2018} TIR intermediate representation to partially unroll the output channel dimension $K$ and input channel dimension $C$ such that the generated instructions can be executed in parallel on the given architecture resulting in a weight stationary dataflow. By unrolling different dimensions of a DNN layer, different dataflows, and data reuse strategies can be explored. However, this is beyond the scope of this paper. The unrolling factors for $K$ and $C$ depend on hardware parameters such as the dimensions of a systolic array instance and the memory port width, which are extracted from the ACADL object diagram. For a complete computation of a convolutional layer, this loop kernel needs to be executed $k$ iteration times with different memory addresses. This $k$ is obtained from the loop variables of the unrolled TIR computation.

Moving to more abstract hardware models, the instructions to be executed also become more abstract. For example, some accelerator architectures only support tiled matrix-matrix multiplications (GEMM) with sometimes fused pooling and activation operations. Therefore, the instructions for an end-to-end latency estimation of a DNN layer need to be generated accordingly. Firstly, by converting the convolutional or fully-connected layer to a GEMM computation using an im2col transformation and, secondly, by tiling the resulting matrix-matrix multiplication to fit the supported tile size of the accelerator.

Overall, to create a valid DNN layer mapping for an architecture modeled in ACADL, one needs to consider the operations that can be executed on the given hardware and then partition the DNN layers according to these operations. For low-level models, this could be \texttt{load} and \texttt{mac} instructions, for intermediate-level models \texttt{gemm} instructions, and for high-level models \texttt{conv\_ext} instructions. 

Lastly, layer fusion, pruning, and quantization are considered differently on the different abstraction levels. For low-level instructions, a tool like TVM can provide an already optimized computation graph where e.g. the activation function is fused into a preceding convolutional layer and unimportant weights and computations have been removed. The unrolling would then be done on the optimized layer. On a higher abstraction level, layer fusion, pruning, and quantization depends on the architecture's capability. In the case of UltraTrail, the quantized and fused \texttt{conv\_ext} layer computes the convolution, pooling, and activation using its MAC-array and output processing unit (OPU). For the systolic array a pruned layer would contain less instructions which could lead to an sub-optimal utilization. As a result, layer fusion, pruning, and quantization is incorporated either explicitly by mapping already optimized layers or implicitly via tensor-level operations.

\section{Architectural Instruction Dependency Graph}
To accurately estimate the performance of DNNs mapped onto an accelerator architecture without using a simulator, we propose the Architectural Instruction Dependency Graph (AIDG) \cite{acadl2022}, capturing when an instruction occupies a hardware module (ACADL object), in which order an instruction propagates through an accelerator (forward), resource conflicts (structural dependencies), data dependencies, and buffer fill levels, extending the Execution Graph proposed by Li et al. \cite{li2006}.

An $AIDG = (N, E)$ is a directed acyclic graph. Nodes in $N = \{(i,o) : (i,o) \in I \times O\}$ of an AIDG represent that an instruction $i \in I$ occupies an ACADL object $o \in O$. There are four different dependency types $D \in \{f,s,d,b\}$ an edge $e \in E = \{(x,y)_\delta : (x,y)_\delta \in N \times N \times D\}$ can represent: forward $f$, structural dependency $s$, data dependency $d$, and a buffer fill level dependency $b$. 

\subsection{Construction}
\label{sec:aidg_construction}

To construct an AIDG from a given set of instructions $I$ and an ACADL object diagram $O$ each instruction $i \in I$ is propagated in order through the ACADL object diagram $O$, which returns the order $\vec{o}(i)$ of ACADL objects $i$ passes through. If an instruction reads from memory, an ACADL object named writeBack is added to the order $\vec{o}(i)$, which is later used as data dependency for nodes accessing the write register of the read operation. For each $o \in \vec{o}(i)$, a node $(i,o)$ is created. The nodes containing $i$ and $o_0, \dots, o_n \in \vec{o}(i)$ are connected according to the order in $\vec{o}(i)$ with forward edges $((i,o_j),(i,o_{j+1}))_f$. Nodes containing the InstructionFetchStage object of consecutive instructions are connected through a buffer fill level dependency edge $((i_j, \text{InstructionFetchStage}), (i_{j+1},\text{InstructionFetchStage}))_b$ which is used to track the fill level of the issue buffer to facilitate issuing multiple instructions at once. Except for the node containing the writeBack object, each node is connected with a structural dependency edge to the node that has previously contained the current ACADL object $o$, which is the last structure user $s_u(o) \in N$, afterward the last structure user is set to the target of the added structural edge. If a FunctionalUnit $o$ has sibling FunctionalUnits $F_s \subset O$, these are FunctionalUnits contained by the same ExecuteStage, the last structure user for all $f_s \in F_s$ is set to $o$, which restricts ExecuteStages from processing multiple instructions at the same time using different FunctionalUnits. For the node $(i, o)$ where $o$ is a FunctionalUnit, data dependency edges for all read and write registers of $i$ are added, coming from the last node that accessed those registers. Afterward, $(i, o)$ is set as the last register reader and writer for the read and write registers of $i$. If $o$ of $(i,o)$ is a Memory, data dependency edges for read and write addresses are added coming from the last node accessing those addresses. The node containing the Memory $o$ is set as the last reader and writer for the addresses. The node containing the writeBack object is then set as the last register writer for all write registers of $i$, as mentioned above. The last step of the AIDG construction algorithm is to merge the nodes of consecutive instruction blocks of size \texttt{port\_width} of the instruction Memory. All consecutive \texttt{port\_width} nodes containing the InstructionMemoryAccessUnit and instruction Memory are merged along their respective structural dependency and buffer fill level edges into a single node. The construction of an AIDG is in $\mathcal{O}(|I|\cdot\vec{o}_\text{max})$ where $\vec{o}_\text{max}$ is the longest path an instruction can take through an ACADL object diagram.

Fig.~\ref{fig:aidg_example} shows the constructed and evaluated AIDG using the ACADL object diagram for the 2$\times$2 systolic array presented in Fig.~\ref{fig:systolic_array_acadl_object_diagram} and the two iterations shown in Fig.~\ref{fig:systolic_array_block_diagram}. Because the instruction Memory $\texttt{port\_width} = 2$ the nodes containing the InstructionMemoryAccessUnit and the instruction Memory are merged for every pair of consecutive instructions. 

\begin{algorithm}[htbp]
\caption{AIDG Evaluation}
\label{algo:aidg_evaluation}
\footnotesize{
\begin{algorithmic}[1]
	\Require $AIDG = (N,E)$
	\State $t_{\text{enter}}: N \to \mathbb{Z}^+_0$ \Comment{node enter time hashmap}
	\State $t_{\text{leave}}: N \to \mathbb{Z}^+_0$ \Comment{node leave time hashmap}
	\State $b_{\text{enter}}: \mathbb{Z}^+_0 \to \mathbb{Z}^+_0$ \Comment{buffer fill level hashmap for $t_{\text{enter}}$}
	\State $b_{\text{forward}}: \mathbb{Z}^+_0 \to \mathbb{Z}^+_0$ \Comment{buffer fill level hashmap for forwarding}
	\State $b_{\text{max}} \gets \text{InstructionFetchStage}.\texttt{issue\_buffer\_size}$
	\State $p \gets \texttt{port\_width}$ of instruction Memory 
	\State $\vec{n} \gets \text{topological node order of } AIDG$
	\ForAll{$n \in \vec{n}$}
		\State $e_{s,\text{in}} \gets (n_{s,\text{in}}, n)_f$
		\Comment{in-going structural edge of $n$}
		\State $N_{d,\text{in}} \gets \{n_d : n_d \in (n_d,n)_d\}$ 
		\Comment{nodes that have out-going data dependency edges to $n$}
		\State $e_{f,\text{in}} \gets (n_{f,\text{in}}, n)_f$, $o_{f,\text{in}} \in n_{f,\text{in}}$
		\Comment{in-going forward edge of $n$}
		\State $e_{b,\text{in}} \gets (n_{b,\text{in}}, n)_b$
		\Comment{in-going buffer fill level dependency edge of $n$}
		\State $N_{f,\text{out}} \gets \{n_{f,\text{out}} : n_{f,\text{out}} \in (n,n_{f,\text{out}})_f\}$ 
		\Comment{nodes that have in-going forward edges from $n$}
		\State $l \gets o.\texttt{latency}$
		\Comment{latency of ACADL object $o$ contained in $n$}

		\LComment{set $t_{\text{enter}}(n)$}
		\If{$e_{s,\text{in}} = \emptyset \wedge e_{f,\text{in}} = \emptyset \wedge e_{b,\text{in}} = \emptyset$} 
			\Comment{$n$ has no in-going edges}
			\State $t_{\text{enter}}(n) \gets 0$

		\ElsIf{$e_{s,\text{in}} \neq \emptyset \wedge e_{f,\text{in}} = \emptyset \wedge e_{b,\text{in}} = \emptyset$} 
			\Comment{$n$ has an in-going structural edge}
			\State $t_{\text{enter}}(n) \gets t_{\text{leave}}(n_{s,\text{in}})$

		\ElsIf{$e_{s,\text{in}} = \emptyset \wedge e_{f,\text{in}} \neq \emptyset \wedge e_{b,\text{in}} = \emptyset$} 
			\Comment{$n$ has an in-going forward edge}
			\State $t_{\text{enter}}(n) \gets t_{\text{leave}}(n_{f,\text{in}})$

		\ElsIf{$e_{s,\text{in}} \neq \emptyset \wedge e_{f,\text{in}} \neq \emptyset \wedge e_{b,\text{in}} = \emptyset$} 
			\Comment{$n$ has an in-going forward and an in-going structural edge}
			\State $t_{\text{enter}}(n) \gets \max(t_{\text{leave}}(n_{s,\text{in}}),t_{\text{leave}}(n_{f,\text{in}}))$

		\ElsIf{$e_{s,\text{in}} \neq \emptyset \wedge e_{f,\text{in}} \neq \emptyset \wedge e_{b,\text{in}} = \emptyset$} 
			\Comment{$n$ has an in-going forward and an in-going buffer edge}
			\State $t_{b} \gets \text{get minimal $t \geq t_{\text{leave}}(n_{f,\text{in}})$ with $b_{\text{enter}}(t) < b_{max}$}$
			\State $t_{\text{enter}}(n) \gets \max(t_{b},t_{\text{leave}}(n_{f,\text{in}}))$
			\State $b_{\text{enter}}(t_b) \gets b_{\text{enter}}(t_b) + 1$
		\EndIf

		\LComment{set $t_{\text{leave}}(n)$}
		\State $t_{\text{stop}} \gets \max(t_{\text{enter}}(n),t_{\text{leave}}(N_{d,\text{in}})) + l$
		\If{$N_{f,\text{out}} = \emptyset$} 
			\Comment{$n$ has no out-going forward edges}
			\State $t_{\text{leave}}(n) \gets t_{\text{stop}}$
		
		\ElsIf{$|N_{f,\text{out}}| = 1 \wedge N_{d,\text{in}} \neq \emptyset$} 
			\Comment{$n$ has exactly one out-going forward edge}
			\State $e_{f,\text{out}} \gets (n, n_{f,\text{out}})_f$
			\Comment{in-going structural dependency edge of $n_{f,\text{out}}$}
			\State $e_{f,\text{out},s,\text{in}} \gets (n_{f,\text{out},s,\text{in}}, n_{f,\text{out}})_s$
			\State $t_{\text{leave}}(n) \gets \max(t_{\text{stop}}, t_{\text{leave}}(n_{f,\text{out},s,\text{in}}))$
		\ElsIf{$|N_{f,\text{out}}| > 1 \wedge N_{d,\text{in}} = \emptyset$} 
			\LComment{$n$ has multiple out-going forward edges and no in-going data dependency edges ($o$ is the InstructionFetchStage.\texttt{issue\_buffer})}
			\State $t_{\text{stop}} \gets t_{\text{enter}} + l$
			\For{$j = 1, \dots, p$}
				\State $t_b \gets \text{get minimal $t \geq t_{\text{stop}}$ with $b_{\text{forward}}(t) < b_{\text{max}}$}$
				\State $t_{\text{leave}} \gets \max(t_{\text{stop}}, t_b)$ 
				\State $b_{\text{forward}}(t_b) \gets b_{\text{forward}}(t_b) + 1$
			\EndFor
		\EndIf
	\EndFor
	\State \Return $t_{\text{enter}}, t_{\text{leave}}, b_\text{enter}, b_\text{forward}$
\end{algorithmic}
}
\end{algorithm}

\subsection{Evaluation}
\label{sec:aidg_evaluation}

\begin{figure}[htbp]
	\includegraphics[width=1.0\linewidth]{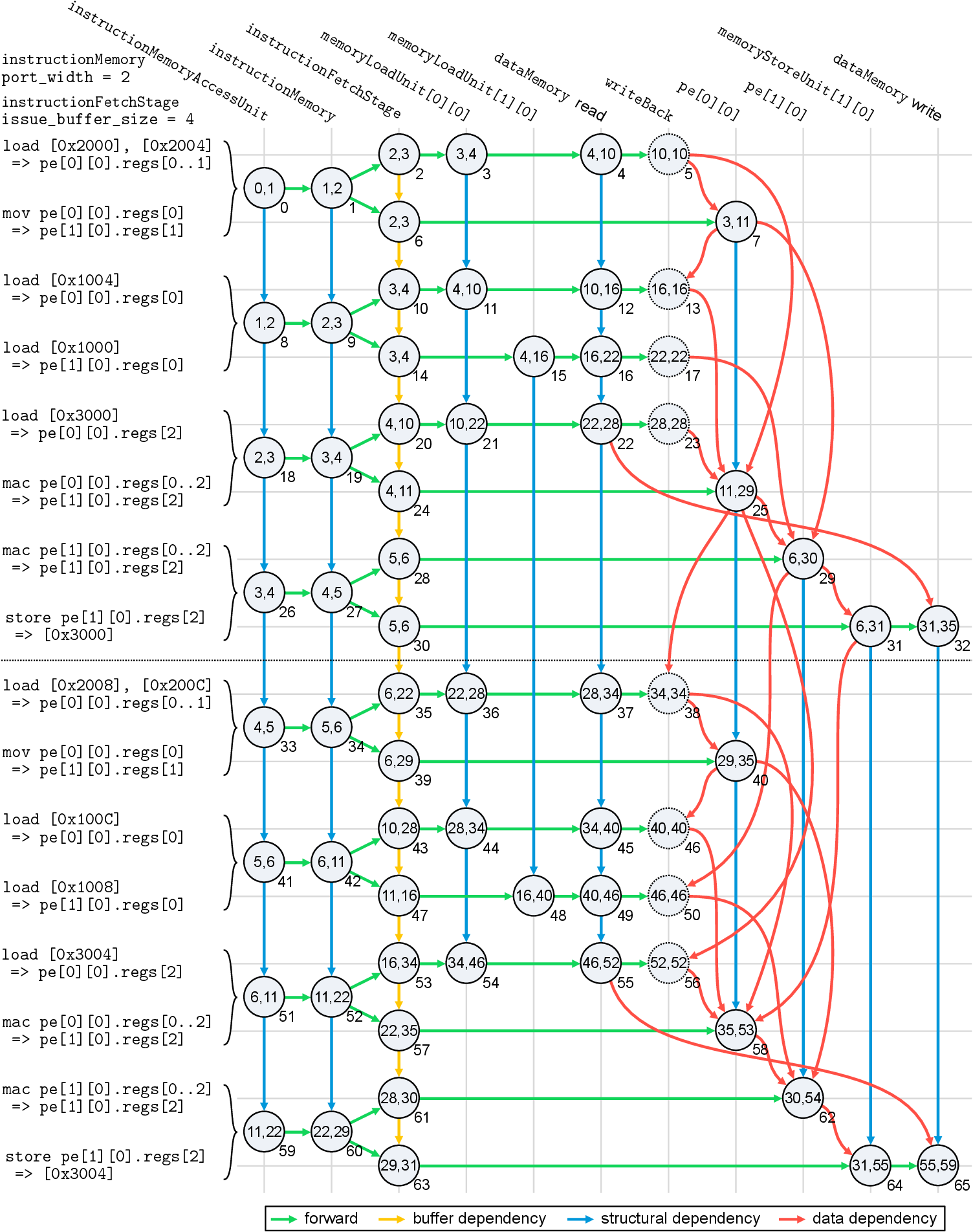}
	\caption{Example of an evaluated AIDG for the 2$\times$2 systolic array. Each node is annotated with the topological order (bottom right of nodes), $t_{\text{enter}}(n)$, and $t_{\text{leave}}(n)$ (inside nodes).}
	\label{fig:aidg_example}
\end{figure}

To determine the end-to-end latency of an AIDG, the nodes are sorted in their topological order. This is done by following the out-going forward edges of each node and, in the instruction order, following the structural dependency edges. Because each node, structural dependency edge, and forward edge is only visited once, computing this topological order is in $\mathcal{O}(|N|+|E_{f,s}|)$. In Fig.~\ref{fig:aidg_example}, the topological order is denoted in the bottom right of each node. For each node, the in-going and out-going edges are evaluated and depending on the different dependency types, $t_{\text{enter}}$ and $t_{\text{leave}}$ are set.  Algorithm \ref{algo:aidg_evaluation} describes in detail how $t_{\text{enter}}$ and $t_{\text{leave}}$ are set for all nodes of an AIDG depending on the types of in-going and out-going edges. 

As an example for determining $t_{\text{enter}}$ and $t_{\text{leave}}$ we pick the nodes $n_{63}$, $n_{64}$, and $n_{65}$ of the topological order in Fig.~\ref{fig:aidg_example}. The \texttt{store} instruction is forwarded from \texttt{instruction\-Memory} to the \texttt{instruction\-FetchStage} at\vspace{-4mm}
\begin{align*}
	t_\text{enter}(n_{63}) = 29.
\end{align*}
Since $n_{63}$ does not have in-going data dependency edges, processing the \texttt{store} instruction by the \texttt{instruction\-FetchStage} starts at $t_\text{enter}(n_{63}) = 29$. At  
\begin{align*}
	t_\text{stop}(n_{63}) = t_\text{enter}(n_{63}) + \texttt{instructionFetchStage.latency} = 29 + 1 = 30
\end{align*}
processing the instruction by the \texttt{instruction\-FetchStage} is finished. Afterward the instruction is forwarded to $n_{64}$ which is connected with an out-going forward edge coming from $n_{63}$. Because $n_{64}$ has an in-going structural dependency edge coming from $n_{31}$, which expresses that the \texttt{memory\-Store\-Unit[1][0]} is occupied by an instruction until $t_\text{leave}(n_{31}) = 31$, the earliest $t_\text{enter}(n_{64})$ is in cycle 31, resulting in stalling the \texttt{store} instruction in the \texttt{instruction\-FetchStage} in $n_{63}$. Therefore,
\begin{align*}
	t_\text{leave}(n_{63}) = \max(t_\text{stop}(n_{63}), t_\text{leave}(n_{31})) = \max(30, 31) = 31.
\end{align*}
After stalling the \texttt{store} instruction in $n_{63}$ the \texttt{memory\-Store\-Unit[1][0]} is no longer occupied and the instruction is forwarded to $n_{64}$ at  
\begin{align*}
	t_\text{enter}(n_{64}) = t_\text{leave}(n_{63}) = 31.
\end{align*}
$n_{64}$ has an in-going data dependency edge coming from $n_{62}$ which expresses that processing the \texttt{store} instruction in the \texttt{memory\-Store\-Unit[1][0]} is stalled until $n_{62}$ has finished processing the \texttt{mac} instruction resulting in
\begin{align*}
	t_\text{stop}(n_{64}) = \max(t_\text{enter}(n_{64}),t_\text{leave}(n_{62})) + \texttt{memoryStoreUnit[1][0].latency} = 54 + 1 = 55.
\end{align*}
$n_{65}$ has an in-going structural dependency edge coming from $n_{32}$ which has to be considered before forwarding the \texttt{store} instruction, however, $t_\text{stop}(n_{64}) > t_\text{leave}(n_{32})$ meaning that the \texttt{store} instruction is not stalled after it finished processing in the \texttt{memory\-Store\-Unit[1][0]}. Therefore, the \texttt{store} instruction is forwarded at
\begin{align*}
	t_\text{leave}(n_{64}) = \max(t_\text{stop}(n_{64}), t_\text{leave}(n_{32})) = \max(55, 35) = 55.
\end{align*}
$n_{65}$ has an in-going data dependency edge coming from $n_{55}$, however, $t_\text{enter}(n_{65}) > t_\text{leave}(n_{55})$ therefore, 
\begin{align*}
	t_\text{stop}(n_{65}) = \max(t_\text{enter}(n_{65}),t_\text{leave}(n_{55})) + \texttt{dataMemory.write\_latency} = 55 + 4 = 59. 
\end{align*}
Since $n_{65}$ has no out-going forward edge 
\begin{align*}
	t_\text{leave}(n_{65}) = t_\text{stop}(n_{65}) =  59.
\end{align*}

Fig.~\ref{fig:aidg_example} shows that data-independent instructions processed by different ACADL objects run in parallel. In contrast, instructions that depend on data from instructions before them are stalled until the input data has been generated. This implicitly facilitates token synchronization via data dependencies over registers or memories and ensures correct calculations even though instructions are issued out-of-order.

The end-to-end latency of an AIDG is
\begin{equation}
	\Delta t_{\text{AIDG}} = \max_{n\in N}(t_{\text{leave}}(n))-\min_{n\in N}(t_{\text{enter}}(n))
\end{equation}
which is the difference between $t_{\text{enter}}$ of the first node in the topological order and the maximum $t_{\text{leave}}$ of all nodes in the AIDG. Considering the AIDG presented in Fig.~\ref{fig:aidg_example} 
\begin{align*}
	\Delta t_{\text{AIDG}} = t_\text{leave}(n_{65}) - t_\text{enter}(n_0) = 59 - 0 = 59.
\end{align*}

The evaluation of an AIDG is in $\mathcal{O}(|N|)$ since each node is only visited once. 

Given an arbitrary ACADL model of an accelerator architecture and instructions on the same abstraction level, we can construct and evaluate an AIDG to make an end-to-end latency estimation. For fused, pruned, and/or quantized DNN layers, the loop kernel to be evaluated in an AIDG would contain fewer and/or different instructions than an unoptimized DNN layer. However, the AIDG construction and evaluation are agnostic towards a loop kernel's number and kind of instructions.

\subsection{End-to-end Latency of a Deep Neural Network Layer}
When executing consecutive iterations in a pipeline, the end-to-end latency of a single iteration cannot be multiplied to get the execution time of the whole loop. This is because successive iterations overlap each other in a pipeline. Moreover, loop-carried dependencies can lead to different end-to-end latencies in consecutive iterations. However, this difference in end-to-end latencies is only present in the first few iterations, while the end-to-end latencies of all following iterations only vary by a negligible amount.

\begin{figure}[htbp]
	\centerline{\includegraphics[width=1.0\linewidth]{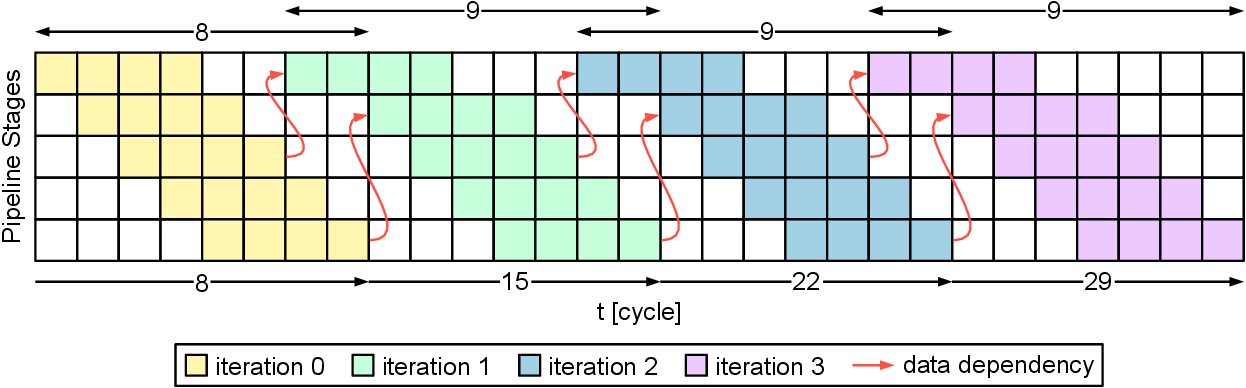}}
	\caption{End-to-end latencies and data dependencies of consecutive iterations executed in a pipeline.}
	\label{pipeline_loops}
\end{figure}

Fig.~\ref{pipeline_loops} shows the overlap of consecutive iterations in a pipeline with different end-to-end latencies of successive iterations of the same loop kernel caused by loop-carried dependencies. While iteration 0 has an execution time of 8 cycles, because there are no dependencies from previous iterations, all following iterations have an end-to-end latency of $\Delta t_{\text{iteration}}=9$. Multiplying the executing time of the first iteration by the number of iterations ($k=4$) results in 32 cycles, while the actual end-to-end latency for the whole loop is 29. After iteration 1, the end-to-end latency of the iterations becomes stable. We call the iterations before the execution times of successive iterations stabilize prolog ($k_{\text{prolog}}=2$, $\Delta t_{\text{prolog}}=15$). The overlap between two successive iterations after the prolog is $\Delta t_{\text{overlap}} = 2$. Considering the prolog and overlap results in the following equation describing the end-to-end latency of a whole loop:
\begin{equation}
	\Delta t = \Delta t_{\text{prolog}} + (k-k_{\text{prolog}})\cdot(\Delta t_{\text{iteration}}-\Delta t_{\text{overlap}}).\label{eqn:delta_t}
\end{equation}

Due to the dataflow-driven nature of DNNs, the loop kernel of a DNN layer does not contain any control-flow instructions. This means each iteration executes the same instructions with different memory addresses. We assume the memory latency of consecutive iterations does not change since accelerators heavily rely on SRAM or standard cell memory for their on-chip memory \cite{ultratrail2020, plasticine2017, eyerissv12017, yodann2016, dianao2014} and because the memory access patterns are always the same. Under this assumption, we can use an AIDG containing at least $k_\text{prolog}$ iterations to determine $\Delta t_{\text{prolog}}$, $\Delta t_{\text{iteration}}$, $\Delta t_{\text{overlap}}$, and $\Delta t$.

To determine $k_\text{prolog}$, we use a fixed point computation as proposed in \cite{li2006} to find the number of iterations until $\Delta t_\text{iterations}$ becomes stable. Firstly, we calculate the minimal number of iterations $k_\text{block}$ such that the number of instructions $|I|$ in $k_\text{block}$ iterations is divisible by the \texttt{port\_width} of the instruction Memory $p$:
\begin{equation}
	k_\text{block} = \frac{\text{lcm}(|I|,p)}{|I|}.
\end{equation}
This is necessary because if an AIDG is constructed consisting of less than $k_\text{block}$ iterations, the merged nodes of the last instructions do not contain $p$ instructions, which can lead to a wrong $\Delta t_\text{iteration}$ and $\Delta t_\text{overlap}$.

We then check if $k_\text{block} \geq k$. If this is the case, the size of $k_\text{prolog}$, which is a multiple of $k_\text{block}$, is longer or equal to the number of iterations $k$ of the whole DNN layer. Therefore, an AIDG for all iterations of the layer is constructed and evaluated, while the end-to-end latency of the layer is $\Delta t_\text{AIDG}$. The same is done if $3\cdot k_\text{block} > k$ because there are insufficient iterations for a fixed point evaluation. Because the first iteration of the first $k_\text{block}$ iterations does not have any in-going structural dependencies, it has a $\Delta t_\text{iteration}$ that is not representative for the whole DNN layer. The next two $k_\text{block}$ iterations have in-going structural dependencies from the previous $k_\text{block}$ iterations and can be used to check for a fixed point. This results in at least three $k_\text{block}$ iterations to apply a fixed point evaluation. 

However, if $3\cdot k_\text{block} \leq k$, an AIDG containing $k_\text{block}$ iterations is constructed and evaluated. From this evaluation, we get
\begin{equation}
	\Delta t_\text{iteration}(k_\text{block}) = \max_{n\in N_{k_\text{block}}}(t_\text{leave}(n)) - \min_{n\in N_{k_\text{block}}}(t_\text{enter}(n))
\end{equation}
which is the end-to-end latency of the last iteration in the AIDG while $N_{k_\text{block}}$ denotes the set of AIDG nodes of the iteration $k_\text{block}$. We then append AIDGs to the first AIDG for the next $k_\text{block}$ iterations until 
\begin{equation}
	\Delta t_\text{iteration}((j-1)\cdot k_\text{block}) = \Delta t_\text{iteration}(j\cdot k_\text{block}).\label{eqn:fixed_point_criterion}
\end{equation}
At this point; we assume that end-to-end latency of consecutive iterations has stabilized. From the last iteration of the last appended AIDG, we determine
\begin{equation}
	k_\text{prolog} = j\cdot k_\text{block},
\end{equation}
\begin{equation}
	\Delta t_\text{iteration} = \max_{n\in N_{k_\text{prolog}}}(t_\text{leave}(n)) - \min_{n\in N_{k_\text{prolog}}}(t_\text{enter}(n)),
\end{equation}
and
\begin{equation}
	\Delta t_\text{overlap} = \max_{n\in N_{k_\text{prolog}}}(t_\text{leave}(n)) - t_\text{enter}((i_\text{last},o_0\in\vec{o}(i_\text{last}))
\end{equation}
Using equation (\ref{eqn:delta_t}) we can now estimate the end-to-end latency $\Delta \hat{t}$ of a whole DNN layer. However, it might be that equation (\ref{eqn:fixed_point_criterion}) is never satisfied due to an oscillating $\Delta t_\text{iteration}$. 

In the case of an oscillating $\Delta t_\text{iteration}$, where equation (\ref{eqn:fixed_point_criterion}) is not satisfied after evaluating 1\% of all iterations $k$, we use a fallback heuristic by constructing an AIDG containing
\begin{equation}
	k_\text{0.01} = \lfloor k\cdot 0.01 \rfloor
\end{equation}
iterations (1\% of all iterations $k$) and set\vspace{-2mm}
\begin{equation}
	k_\text{prolog} = \left\lfloor\frac{k_\text{0.01}}{4}\right\rfloor.
\end{equation}
Afterward, we determine
\begin{equation}
	\Delta t_{k_\text{prolog},k_\text{0.01}} = \max_{n\in N_{k_\text{0.01}}}(t_\text{leave}(n)) - \max_{n\in N_{k_\text{prolog}}}(t_\text{leave}(n))
\end{equation}
which is the latency from the last iteration $k_\text{prolog}$ in the prolog to the last iteration $k_\text{0.01}$ in the AIDG. Using this AIDG we approximate
\begin{equation}
	\Delta t_\text{iteration} = \frac{\Delta t_{k_\text{prolog},k_\text{0.01}}}{k_\text{0.01}-k_\text{prolog}}
\end{equation}
and set\vspace{-2mm}
\begin{equation}
	\Delta t_\text{overlap} = 0.
\end{equation}
Now we can use (\ref{eqn:delta_t}) to estimate the end-to-end latency $\Delta \hat{t}$.

The number of iterations with which an AIDG for the fallback heuristic is constructed was empirically determined by evaluating different architectures and mappings with increasing percentages of $k$. Constructing and evaluating an AIDG containing $1\%$ of all iterations offers a good trade-off between estimation runtime and estimation accuracy (see Appendix \ref{sec:parameter_determination}).

\section{Performance Model Evaluation and Results}
After detailing our proposed performance estimation approach for DNN accelerators, this section presents and discusses the obtained modeling and estimation results for four different accelerator architectures and mapping the DNNs TC-ResNet8 \cite{tcresnet2019}, EfficientNet \cite{efficientnet2019}, and AlexNet \cite{alexnet2017} onto them. Those DNNs cover typical edge AI use-cases with the following layer types: 1D-, 2D-, and depth-wise convolution; fully-connected; average and max pooling; ReLU and clip activation; element-wise addition and multiplication; together with residual connections, which provide a wide variety of different workloads.

As a comparison against regression models for DNN latency estimations, we use the best MAPE of 7.67\si{\percent} of a support vector regression reported by Bouzidi et al. \cite{bouzidi2021}. We did not create regression models ourselves because large datasets are needed to achieve reasonable accuracy. For example, Bouzidi et al. \cite{bouzidi2021} conducted at least \num{200000} measurements per platform to create a dataset to train their estimators. Since a single measurement using the Verilator RTL simulator for the Gemmini accelerator takes almost nine minutes for the smallest DNN TC-ResNet, generating only \num{10000} samples would take two months. Therefore, we can not report on the runtime, estimated cycles, and percentage error for regression models and only report the MAPE. Additionally, we used the refined roofline model reported by Wess et al. \cite{annette2021} to compare our estimation approach with an analytical model. While the traditional roofline model \cite{roofline2009} only considers the peak compute performance and peak memory bandwidth, the refined roofline model also incorporates the unrolling parameters of a DNN layer. We also created a Timeloop \cite{timeloop2019} model for the Gemmini accelerator, which is also an analytical performance model.

We compare the different estimators with ground-truth values gathered from simulators using the estimated cycles for a whole DNN 
\vspace{-3mm}
\begin{equation}
	\hat{T} = \sum^n_{i=1} \Delta \hat{t}_i.
\end{equation}
Furthermore, we report the accuracy of the compared estimators using the percentage error
\begin{equation}
	\text{PE} = \frac{\hat{T} - \sum^{n}_{i=1}\Delta t_i}{\sum^{n}_{i=1}\Delta t_i} \cdot 100\%
\end{equation}
for a whole DNN. Additionally, we use the mean absolute percentage error
\begin{equation}
	\text{MAPE} = \frac{1}{n} \sum^{n}_{i=1}\left|\frac{\Delta t_i - \Delta \hat{t}_i}{\Delta t_i}\right| \cdot 100\%
\end{equation}
as an end-to-end latency estimation accuracy metric for layers of a whole DNN. We also report the runtime of our end-to-end latency estimation approach. All runtime measurements were conducted on an Intel Xeon Platinum 8168 CPU with 192\,GB RAM running Ubuntu Linux 20.04 while the AIDG construction and evaluation is implemented as a single-threaded application.

\subsection{UltraTrail}
For the UltraTrail accelerator, we created an ACADL model on the tensor operations level, as introduced in section \ref{sec:acadl_modeling_examples}, and performed an AIDG evaluation. Moreover, we built a refined roofline model for UltraTrail. Table \ref{tab:ultratrail_estimator_comparison} compares different latency estimators with the cycle-accurate Cadence Xcelium RTL simulator when mapping TC-ResNet8 onto UltraTrail. Since UltraTrail only supports one-dimensional data processing and convolutional layers \cite{ultratrail2020} the DNNs EfficientNet and AlexNet can not run on this accelerator architecture because they rely heavily on two-dimensional convolutional layers. Therefore, we only report results for the TC-ResNet8, which consists of one-dimensional convolutional and fully-connected layers. Our AIDG-based end-to-end latency estimation outperforms the refined roofline model, and the literature-reported regression model regarding PE and MAPE matching the ground truth almost precisely. Additionally, the estimation runtime of our approach on this abstraction level is comparable to analytical models and allows for fast and accurate performance estimations. The peak memory usage for estimating the end-to-end latency of all layers of the TC-ResNet8 using the AIDG evaluation is 146\,MiB.

\begin{table}[htbp]
	\caption{Comparison of latency estimators for the TC-ResNet8 mapped onto the UltraTrail accelerator.}
	\label{tab:ultratrail_estimator_comparison}
	\begin{center}
	\begin{tabular}{lcccc}
	\toprule
	Estimator & Runtime & Estimated cycles $\hat{T}$ & PE & MAPE\\
	\midrule
        AIDG & 22\si{\milli\second} & \num{22484} & \textbf{0.013\si{\percent}} & \textbf{0.0001\si{\percent}}\\
	Regression model \cite{bouzidi2021} & -- & -- & -- & 7.67\si{\percent}\\
        Refined roofline \cite{annette2021} & \textbf{< 1\si{\milli\second}} & \num{24168} & 7.5\si{\percent} & 6.37\si{\percent}\\
	\addlinespace
	\midrule
	RTL simulator & Runtime & Measured cycles\\
	\midrule
        Cadence Xcelium & 2.79\si{\second} (4.28\si{\second}\textsuperscript{1}) & \num{22481} & \multicolumn{2}{c}{ground truth}\\ 
	\bottomrule
	\end{tabular}
        \newline
        \newline
        \scriptsize{\textsuperscript{1} Runtime for non pre-compiled RTL code}
	\end{center}
\end{table}

\subsection{Gemmini}
\label{sec:gemmini_results}
Gemmini \cite{gemmini2021} is a parameterizable accelerator generator developed by UC Berkeley and is part of the Chipyard \cite{chipyard2020} ecosystem. At its core, Gemmini is composed of a systolic array of DIM$\times$DIM MAC units. The accelerator is complemented by a scratchpad of banked SRAMs, used to store the matrix-matrix multiplication (GEMM) inputs $A$, $B$, $D$, and another one equipped with adder units known as the accumulator, used to store the output matrix $C$ of the computation $C=A\cdot B + D$. Additionally, Gemmini can be configured to apply activation and pooling functions to outputs of GEMM operations and thereby supporting on device layer fusion. A DMA engine is used to exchange data between the accelerator's SRAMs and the SoC's L2 cache. Gemmini works in a tightly-coupled manner with a RISC-V CPU by receiving custom instructions through a Rocket Chip Co-processor (RoCC) interface. Internally, Gemmini uses a decoupled access-execute architecture, which allows for accessing data and processing it in parallel. A reorder buffer detects hazards between Gemmini instructions and issues them to their respective controller in the correct order. For our performance model comparisons, we instantiated a Gemmini accelerator with a DIM size of 16.

\begin{figure}[htbp]
	\centerline{\includegraphics[width=1.0\linewidth]{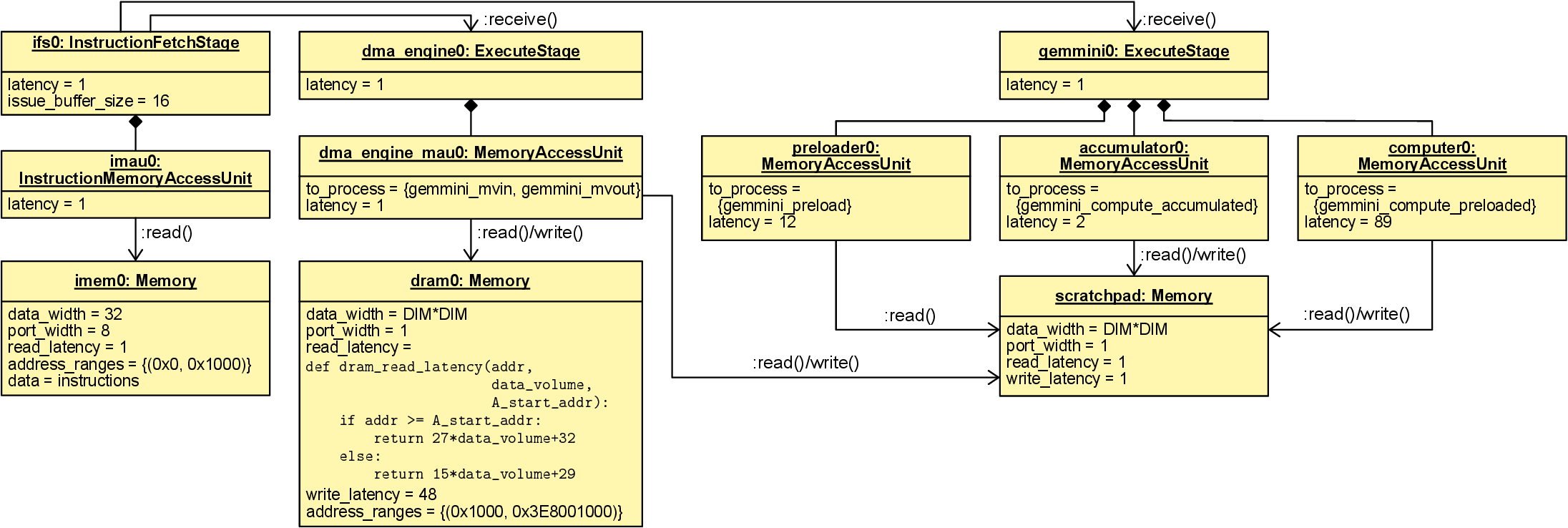}}
	\caption{Gemmini ACADL object diagram.}
	\label{fig:gemmini_acadl_object_diagram}
\end{figure}

Fig.~\ref{fig:gemmini_acadl_object_diagram} depicts the ACADL object diagram for the Gemmini accelerator obtained from the architectural block diagram and benchmarking the \texttt{gemmini\_mvin}, \texttt{gemmini\_preload}, \texttt{gemmini\_compute\-\_accumulated}, etc. instructions that are directly taken from publicly available Gemmini examples. The decoupled access-execute architecture is modeled in ACADL using the two parallel ExecuteStages \texttt{dma\_engine0} and \texttt{gemmini0}. The FunctionalUnits of those two ExecuteStages can independently access the scratchpad and are executed when existing dependencies are resolved, closely modeling Gemmini's reorder buffer. For the \texttt{dram0} read latency we use a simple linear latency model which incorporates the accessed data volume and start address of the matrix $A$ to accomodate for DRAM burst access latencies. 

Using the im2col transformation \cite{im2col2006} we turned the convolutional layers of the TC-ResNet8, AlexNet, and EfficientNet into GEMM operations, which we then split into tiles of 16$\times$16 matching Gemmini's DIM size. The fully-connected layers were also split into 16$\times$16 tiles without the need for a transformation. Using the aforementioned Gemmini instructions, we implemented a tiled GEMM to execute the DNN layers of TC-ResNet8, AlexNet, and EfficientNet on the cycle-accurate Verilator RTL model to obtain a ground truth for the end-to-end latencies. We also used the instructions from those implementations for our AIDG evaluation. Since the implementation is composed of nested loops, we were able to use the AIDG fixed point evaluation, which led to a significant speedup of the estimation runtime. 

To compare our latency estimation approach against analytical models, we used Timeloop \cite{timeloop2019} and created a model of the Gemmini accelerator. Given Timeloop's limitations, due to its coarse textual description, the Gemmini model is limited in its functionality. First, due to Timeloop's lack of support for parallel memory structures in the hierarchy, our model contains a dependence between the tiling of the accumulator and the scratchpad. This dependence does not exist in the real hardware, as scratchpad and accumulator are completely independent memories.

Second, because Timeloop cannot simulate pipeline stalls or structural conflicts, we used the simplex method \cite{simplex1965} using Verilator measurements as input in order to find the best read and write bandwidths for each memory. This helps mitigate this Timeloop limitation, by limiting the actual bandwidth of each memory to encapsulate the effects of the decoupled access-execute structure of Gemmini on the real hardware. As a second analytical model, we also created a refined roofline model for Gemmini based on the model introduced by Wess et al. \cite{annette2021}.  

\begin{table}[htbp]
    \caption{Comparison of latency estimators for TC-ResNet8 mapped onto the 16$\times$16 Gemmini accelerator.}
	\label{tab:gemmini_tc_resnet_estimator_comparison}
	\begin{minipage}{\columnwidth}
	\begin{center}
	\begin{tabular}{lcccc}
	\toprule
	Estimator & Runtime & Estimated cycles $\hat{T}$ & PE & MAPE \\
	\midrule
    AIDG fixed point eval. & 0.5\si{\second} & \num{37384} & \textbf{1.1\si{\percent}} & \textbf{3.67\si{\percent}} \\
	Regression model \cite{bouzidi2021} & -- & -- & -- & 7.67\si{\percent} \\
        Refined roofline \cite{annette2021} & \textbf{< 1\si{\milli\second}} & \num{35943} & -3.0\si{\percent} & 12.80\si{\percent} \\
    Timeloop \cite{timeloop2019} & 12.6\si{\second} & \num{28266} & -23.56\si{\percent} & 28.93\si{\percent} \\
	\addlinespace
	\midrule
	RTL simulator & Runtime & Measured cycles & & \\
	\midrule
        Verilator & \num{527}\si{\second} (8.78\si{\minute}) & \num{36979} & \multicolumn{2}{c}{ground truth} \\ 
	\bottomrule
	\end{tabular}
	\end{center}
	\end{minipage}
\end{table}

\begin{table}[htbp]
    \caption{Comparison of latency estimators for AlexNet mapped onto the 16$\times$16 Gemmini accelerator.}
	\label{tab:gemmini_alexnet_estimator_comparison}
	\begin{minipage}{\columnwidth}
	\begin{center}
	\begin{tabular}{lcccc}
	\toprule
	Estimator & Runtime & Estimated cycles $\hat{T}$ & PE & MAPE\\
	\midrule
        AIDG fixed point eval. & 37.9\si{\second} & \num{35777113} & \textbf{-2.02\si{\percent}} & 9.78\si{\percent} \\
        Regression model \cite{bouzidi2021} & -- & -- & -- & \textbf{7.67\si{\percent}}\\
        Refined roofline \cite{annette2021} & \textbf{< 1\si{\milli\second}} & \num{28772598} & -21.0\si{\percent} & 30.92\si{\percent}\\
	Timeloop \cite{timeloop2019} & 9.5\si{\second} & \num{29251930} & -20.86\si{\percent} & 48.25\si{\percent}\\
	\addlinespace
	\midrule
	RTL simulator & Runtime & Measured cycles \\
	\midrule
        Verilator & \num{156630}\si{\second} (43.5\si{\hour}) & \num{36961948} & \multicolumn{2}{c}{ground truth}\\ 
	\bottomrule
	\end{tabular}
	\end{center}
	\end{minipage}
\end{table}

\begin{table}[htbp]
    \caption{Comparison of latency estimators for EfficientNet mapped onto the 16$\times$16 Gemmini accelerator.}
	\label{tab:gemmini_efficient_net_estimator_comparison}
	\begin{minipage}{\columnwidth}
	\begin{center}
	\begin{tabular}{lcccc}
	\toprule
	Estimator & Runtime & Estimated cycles $\hat{T}$ & PE & MAPE\\
	\midrule
        AIDG fixed point eval. & 17.3\si{\second} & \num{7626267} & \textbf{-0.56\si{\percent}} & \textbf{7.51\si{\percent}}\\
	Regression model \cite{bouzidi2021} & -- & -- & -- & 7.67\si{\percent}\\
        Refined roofline \cite{annette2021} & \textbf{< 1\si{\milli\second}} & \num{7269663} & -5.21\si{\percent} & 21.9\si{\percent}\\
    Timeloop \cite{timeloop2019} & 76.9\si{\second} & \num{7423733} & -3.2\si{\percent} & 14.02\si{\percent}\\
	\addlinespace
	\midrule
	RTL simulator & Runtime & Measured cycles \\
	\midrule
        Verilator & \num{42919}\si{\second} (11.9\si{\hour})& \num{7669519} & \multicolumn{2}{c}{ground truth}\\ 
	\bottomrule
	\end{tabular}
	\end{center}
	\end{minipage}
\end{table}

\begin{figure}[htbp]
	\centerline{\includegraphics[width=\linewidth]{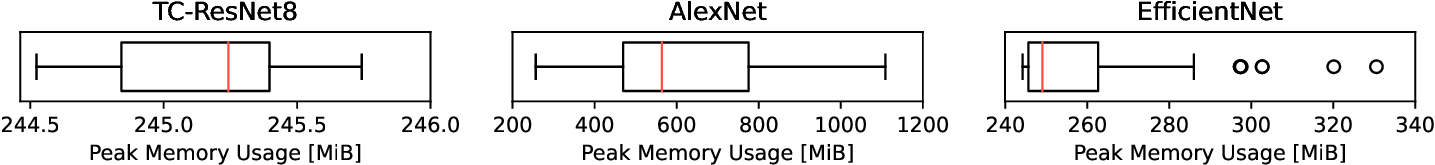}}
	\caption{Peak memory usage of the AIDG fixed point evaluation for TC-ResNet, AlexNet, and EfficientNet mapped onto the 16$\times$16 Gemmini accelerator. The box represents the inter-quartile range from the first to the third quartile, while the line inside represents the median. The whiskers extend from the box by 1.5 times the inter-quartile range. The circles are outliers.}
	\label{fig:gemmini_memory_usage}
\end{figure}

Tables \ref{tab:gemmini_tc_resnet_estimator_comparison}, \ref{tab:gemmini_alexnet_estimator_comparison}, and \ref{tab:gemmini_efficient_net_estimator_comparison} present the obtained results of the different latency estimators for TC-ResNet8, AlexNet, and EfficientNet mapped onto the $16\times 16$ Gemmini accelerator, compared against the cycle-accurate Verilator RTL simulator. Our estimation approach outperforms the refined roofline model and Timeloop regarding MAPE. Both analytical models cannot accurately model Gemmini's decoupled access-execute architecture, leading to a significant deviation from the ground truth. For EfficientNet, Timeloop has the best PE because over- and underestimation balance each other out in this case. Regarding estimation runtime, our proposed approach is several magnitudes faster than an RTL simulation and faster than Timeloop for TC-ResNet8 and EfficientNet. The estimation runtime of Timeloop can largely be attributed to the instantiation of the Accelergy \cite{accelergy2019} backend, which is done separately for each layer and contributes at least 1\si{\second} to the runtime per layer. Fig.~\ref{fig:gemmini_memory_usage} shows box plots for the peak memory usage of the AIDG fixed point evaluation for all layers of TC-ResNet, EfficientNet, and AlexNet. For all three DNNs, the peak memory usage is below 1200\,MiB.

\subsection{Parameterizable Systolic Array}
\label{sec:systolic_array_results}

To validate that our ACADL-based accelerator modeling and AIDG fixed point evaluation also work for parameterizable architectures, we mapped the DNNs TC-ResNet8, AlexNet, and EfficientNet onto different sizes of the systolic array introduced in section \ref{sec:acadl_modeling_examples}. 

We instantiated the systolic array in different sizes, which refers to the number of processing elements. First, we ran an AIDG whole graph evaluation, meaning all iterations were evaluated for each DNN/systolic array pairing. The estimated cycles from this AIDG whole graph evaluation are used as a ground truth for comparison (measured cycles). Second, we performed our AIDG fixed point evaluation and built a refined roofline model for the parameterizable systolic array.

Table~\ref{tab:systolic_array_results} presents the obtained results of our extensive systolic array evaluation. Comparing the AIDG fixed point evaluation with the refined roofline model in terms of PE and MAPE shows that our estimation approach yields very accurate results while maintaining a reasonable runtime. Especially for AlexNet, we see a massive reduction in the number of DNN layer iterations that must be evaluated. In the best case, we only have to evaluate 154 iterations to estimate the latency of 4.19 billion instructions. For the 2$\times$2 systolic array our AIDG fixed point evaluation is able to perfectly match the measured cycles because there are almost no pipeline effects and $\Delta t_\text{iteration}$ stays constant after equation (\ref{eqn:fixed_point_criterion}) is satisfied. The roofline model especially struggels with larger array sizes because it does not accurately capture the increasing pipeline effects of the larger systolic arrays. Those pipeline effects lead to a varying utilization efficiencies while the refined roofline model expects a constant utilization efficiency. For the systolic array sizes 4$\times$4 and 8$\times$8, we see higher PE and MAPE for the AIDG fixed point evaluation of all three DNNs. This can largely be attributed to oscillating $\Delta t_\text{iteration}$ and $\Delta t_\text{overlap}$ which are an artifact of sub-optimal DNN mappings in which an unrolled loop kernel underutilizes the systolic array (see Appendix \ref{sec:premature_convergence}). However, our approach shows better MAPE for all systolic array configurations and mappings than the support vector regression MAPE of 7.67\% reported by Bouzidi et al. \cite{bouzidi2021} without needing an extensive training dataset. Due to space constraints, we did not include a column for the regression-based latency estimator MAPE in Table \ref{tab:systolic_array_results}.

\begin{table}[htbp]
	\caption{AIDG fixed point evaluation vs. refined roofline model for varying systolic array sizes.}
	\label{tab:systolic_array_results}
	\begin{minipage}{\columnwidth}
	\begin{center}
	\tiny{
        {
        \setlength{\tabcolsep}{0.3em}
        \begin{tabular}{clrrrrrcc c rcc r}
            \toprule
            & & & & \multicolumn{5}{c}{AIDG fixed point evaluation} & & \multicolumn{3}{c}{Refined roofline}\\
            \cline{5-9}\cline{11-13}
            \addlinespace
            Size &  DNN & $\sum$ iters. & $\sum$ insts. & Eval. iters & Runtime & Est. cycles $\hat{T}$ & PE & MAPE & & Est. cycles $\hat{T}$ & PE & MAPE & Meas. cycles\\
            \midrule
            \multirow{3}*{2$\times$2} & TC-ResNet8 & \num{396824} & \num{6233280} & \num{154} (\num{0.0388}\si{\percent}) & \num{4}\si{\second} & \num{2724117} & \textbf{\num{0.0}\si{\percent}} & \textbf{\num{0.0}\si{\percent}} &  & \num{2723976} & \num{-0.01}\si{\percent} & \num{0.24}\si{\percent} & \num{2724117} \\
            & AlexNet & \num{281577920} & \num{4192359296} & \num{154} (\num{0.0001}\si{\percent}) & \num{4308}\si{\second} & \num{1866213921} & \textbf{\num{0.0}\si{\percent}} & \textbf{\num{0.0}\si{\percent}} &  & \num{1917022208} & \num{2.72}\si{\percent} & \num{1.12}\si{\percent} & \num{1866213921} \\
            & EfficientNet & \num{325762286} & \num{3325722004} & \num{1940} (\num{0.0006}\si{\percent}) & \num{2182}\si{\second} & \num{1399682509} & \textbf{\num{0.0}\si{\percent}} & \textbf{\num{0.0}\si{\percent}} &  & \num{1369716234} & \num{-2.14}\si{\percent} & \num{7.9}\si{\percent} & \num{1399682509} \\
            \addlinespace
            \multirow{3}*{4$\times$4} & TC-ResNet8 & \num{100364} & \num{6600272} & \num{412} (\num{0.4105}\si{\percent}) & \num{9}\si{\second} & \num{688789} & \textbf{\num{-6.49}\si{\percent}} & \textbf{\num{2.91}\si{\percent}} &  & \num{688500} & \num{-6.53}\si{\percent} & \num{3.68}\si{\percent} & \num{736623} \\
            & AlexNet & \num{66361376} & \num{4323923200} & \num{300} (\num{0.0005}\si{\percent}) & \num{3993}\si{\second} & \num{462922413} & \textbf{\num{-5.85}\si{\percent}} & \textbf{\num{2.59}\si{\percent}} &  & \num{454454144} & \num{-7.57}\si{\percent} & \num{3.42}\si{\percent} & \num{491668085} \\
            & EfficientNet & \num{84438277} & \num{2832419380} & \num{58528} (\num{0.0693}\si{\percent}) & \num{1719}\si{\second} & \num{372183005} & \textbf{\num{-2.78}\si{\percent}} & \textbf{\num{1.73}\si{\percent}} &  & \num{370439101} & \num{-3.23}\si{\percent} & \num{8.78}\si{\percent} & \num{382815952} \\
            \addlinespace
            \multirow{3}*{6$\times$6} & TC-ResNet8 & \num{65600} & \num{7384544} & \num{2279} (\num{3.4741}\si{\percent}) & \num{19}\si{\second} & \num{552453} & \textbf{\num{0.04}\si{\percent}} & \textbf{\num{0.01}\si{\percent}} &  & \num{579968} & \num{5.02}\si{\percent} & \num{5.63}\si{\percent} & \num{552234} \\
            & AlexNet & \num{42020032} & \num{5010642432} & \num{295337} (\num{0.7028}\si{\percent}) & \num{5679}\si{\second} & \num{378815947} & \textbf{\num{0.03}\si{\percent}} & \textbf{\num{0.01}\si{\percent}} &  & \num{379711701} & \num{0.26}\si{\percent} & \num{2.97}\si{\percent} & \num{378712216} \\
            & EfficientNet &  \num{41410130} & \num{2994120996} & \num{86718} (\num{0.2094}\si{\percent}) & \num{2246}\si{\second} & \num{215680130} & \textbf{\num{0.42}\si{\percent}} & \textbf{\num{0.36}\si{\percent}} &  & \num{222742036} & \num{3.71}\si{\percent} & \num{11.66}\si{\percent} & \num{214780694} \\
            \addlinespace
            \multirow{3}*{8$\times$8} & TC-ResNet8 & \num{25864} & \num{9131928} & \num{1795} (\num{6.9401}\si{\percent}) & \num{59}\si{\second} & \num{303289} & \textbf{\num{-6.75}\si{\percent}} & \textbf{\num{2.65}\si{\percent}} &  & \num{366864} & \num{12.79}\si{\percent} & \num{6.98}\si{\percent} & \num{325251} \\
            & AlexNet & \num{18807824} & \num{5960619776} & \num{534} (\num{0.0028}\si{\percent}) & \num{6112}\si{\second} & \num{215417988} & \textbf{\num{-3.37}\si{\percent}} & \textbf{\num{7.1}\si{\percent}} &  & \num{258536064} & \num{15.97}\si{\percent} & \num{9.85}\si{\percent} & \num{222941934} \\
            & EfficientNet & \num{21306120} & \num{3367270728} & \num{42752} (\num{0.2007}\si{\percent}) & \num{2618}\si{\second} & \num{131848959} & \textbf{\num{-2.35}\si{\percent}} & \textbf{\num{1.17}\si{\percent}} &  & \num{144797146} & \num{7.24}\si{\percent} & \num{18.19}\si{\percent} & \num{135022277} \\
            \addlinespace
            \multirow{3}*{16$\times$16} & TC-ResNet8 & \num{8661} & \num{13963232} & \num{1830} (\num{21.1292}\si{\percent}) & \num{306}\si{\second} & \num{208927} & \textbf{\num{-0.85}\si{\percent}} & \textbf{\num{0.31}\si{\percent}} &  & \num{217326} & \num{3.38}\si{\percent} & \num{26.44}\si{\percent} & \num{210216} \\
            & AlexNet & \num{5820296} & \num{9808927488} & \num{2282} (\num{0.0392}\si{\percent}) & \num{14295}\si{\second} & \num{130739295} & \textbf{\num{0.34}\si{\percent}} & \textbf{\num{5.15}\si{\percent}} &  & \num{150242768} & \num{15.31}\si{\percent} & \num{32.96}\si{\percent} & \num{130290084} \\
            & EfficientNet & \num{5943150} & \num{4855568488} & \num{52615} (\num{0.8853}\si{\percent}) & \num{4784}\si{\second} & \num{73134760} & \textbf{\num{-0.14}\si{\percent}} & \textbf{\num{1.62}\si{\percent}} &  & \num{81568166} & \num{11.37}\si{\percent} & \num{37.12}\si{\percent} & \num{73240796} \\
            \bottomrule
        \end{tabular}
        }
	}
	\end{center}
	\end{minipage}
\end{table}

\begin{figure}[htbp]
	\centerline{\includegraphics[width=\linewidth]{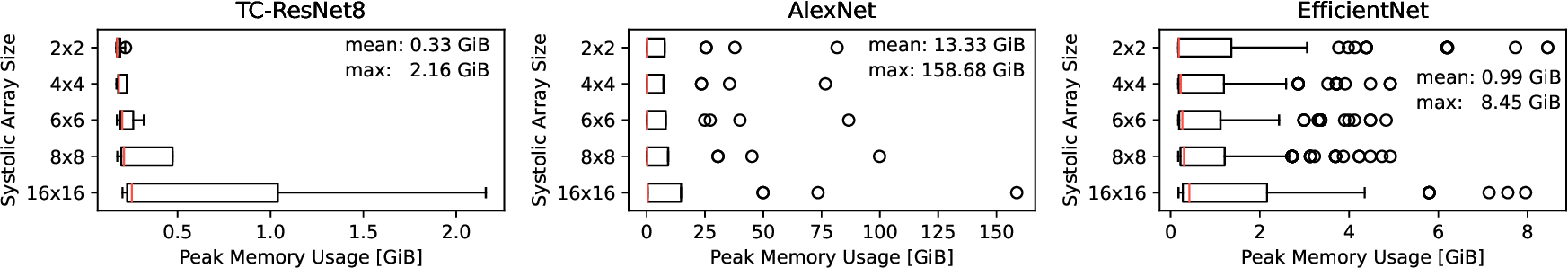}}
	\caption{Peak memory usage of the AIDG fixed point evaluation for varying systolic array sizes and DNNs. The box represents the inter-quartile range from the first to the third quartile, while the line inside represents the median. The whiskers extend from the box by 1.5 times the inter-quartile range. The circles are outliers.}
	\label{fig:systolic_array_memory_usage}
\end{figure}

Fig.~\ref{fig:systolic_array_memory_usage} presents the peak memory usage of the AIDG fixed point evaluation for our systolic array evaluation presented in Table~\ref{tab:systolic_array_results}. Each box plot contains the peak memory usage of all layers of the mapped DNN. For TC-ResNet8, the peak memory usage for all layers and systolic array sizes is below 2.2\,GiB, with a mean of 330\,MiB. The AIDG fixed point evaluation for AlexNet mapped onto the systolic array resulted in a mean memory usage of 13.33\,GiB, with most outliers being under 100\,GiB of peak memory usage. The maximum peak memory usage for AlexNet is at 158.68\,GiB, which resulted from mapping a large convolutional layer, containing 2496 instructions per iteration, onto a 16$\times$16 systolic array and evaluating seven iterations until $\Delta t_\text{iteration}$ becomes stable. For mapping EfficientNet onto the systolic array, the peak memory usage is always below 8.5\,GiB with a mean of 0.99\,GiB. Except for the outliers of the AlexNet mappings, all measurements can be conducted on a modern-day off-the-shelf workstation computer.

To further illustrate the benefits of our proposed approach, we conducted a case study for a systolic array of size 12$\times$12 with varying memory port width. The memory port width is the number of data words that can be accesses in a single memory transaction. 

\begin{figure}[htbp]
	\centerline{\includegraphics[width=1.0\linewidth]{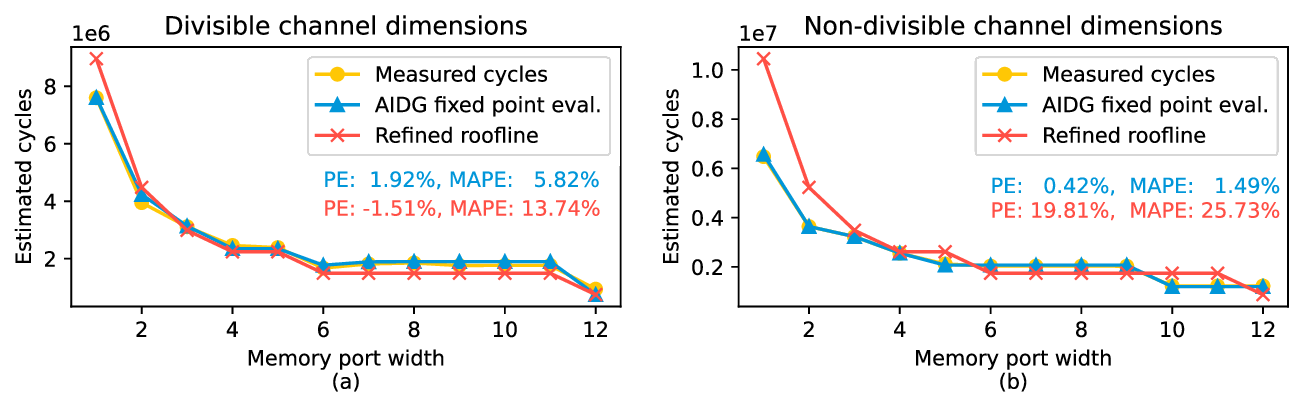}}
	\caption{Estimated cycles of the AIDG fixed point evaluation vs. refined roofline model for $12 \times 12$ systolic array while varying the memory port width.}
	\label{fig:case_study} 
\end{figure}

We mapped two different convolutional layers onto the architecture, one with input and output channel dimensions that are divisible by the size of the systolic array ($C=12$ and $K=72$) and one with non-divisible input and output channel dimensions ($C=20$ and $K=70$). The resulting mapping is, therefore, close-to-optimal in the first case, where all 12 rows and columns of the systolic array can be utilized. In the second case, only $10 \times 10$ MAC operations can be unrolled, resulting in two idle rows and columns. The estimated runtime is depicted in Fig.~\ref{fig:case_study}.

In the case of the divisible convolution, one can clearly see that with a wider memory interface the architecture is less memory bound and the number of estimated cycles reduces. Note that there is no change in runtime between a memory port width of 7 and 11 because for 12 weights that need to be loaded per column we still need two memory transactions with these port widths, which is the same as with a memory port width of 6. 
This behavior is also captured by the roofline model. The accuracy of the roofline estimation compared to the AIDG fixed point evaluation changes in the case of the non-divisible convolution. In Fig.~\ref{fig:case_study}(b), we can see that our approach captures the runtime behavior of the architecture better in case of a sub-optimal mapping and different memory port widths, which is especially important if a performance estimator is to be used in a hardware-aware NAS optimization loop where different DNN layer dimensions and hardware parameters are explored to break out of local minima and detect maxima.

\subsection{Plasticine-derived Accelerator Design Space Exploration}

\begin{figure}[htbp]
	\centerline{\includegraphics[width=1.0\linewidth]{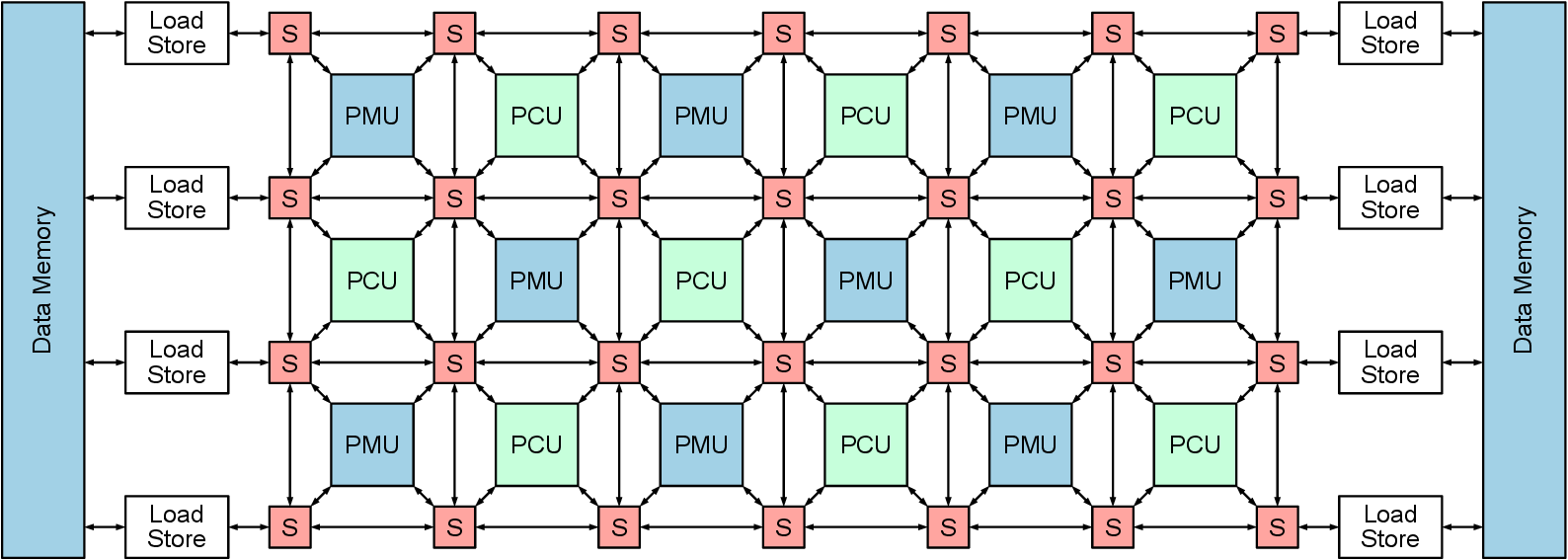}}
	\caption{Block diagram of a 3$\times$6 Plasticine-derived accelerator architecture \cite{plasticine2017}.}
	\label{fig:plasticine_block_diagram}
\end{figure}

\begin{figure}[htbp]
	\centerline{\includegraphics[width=1.0\linewidth]{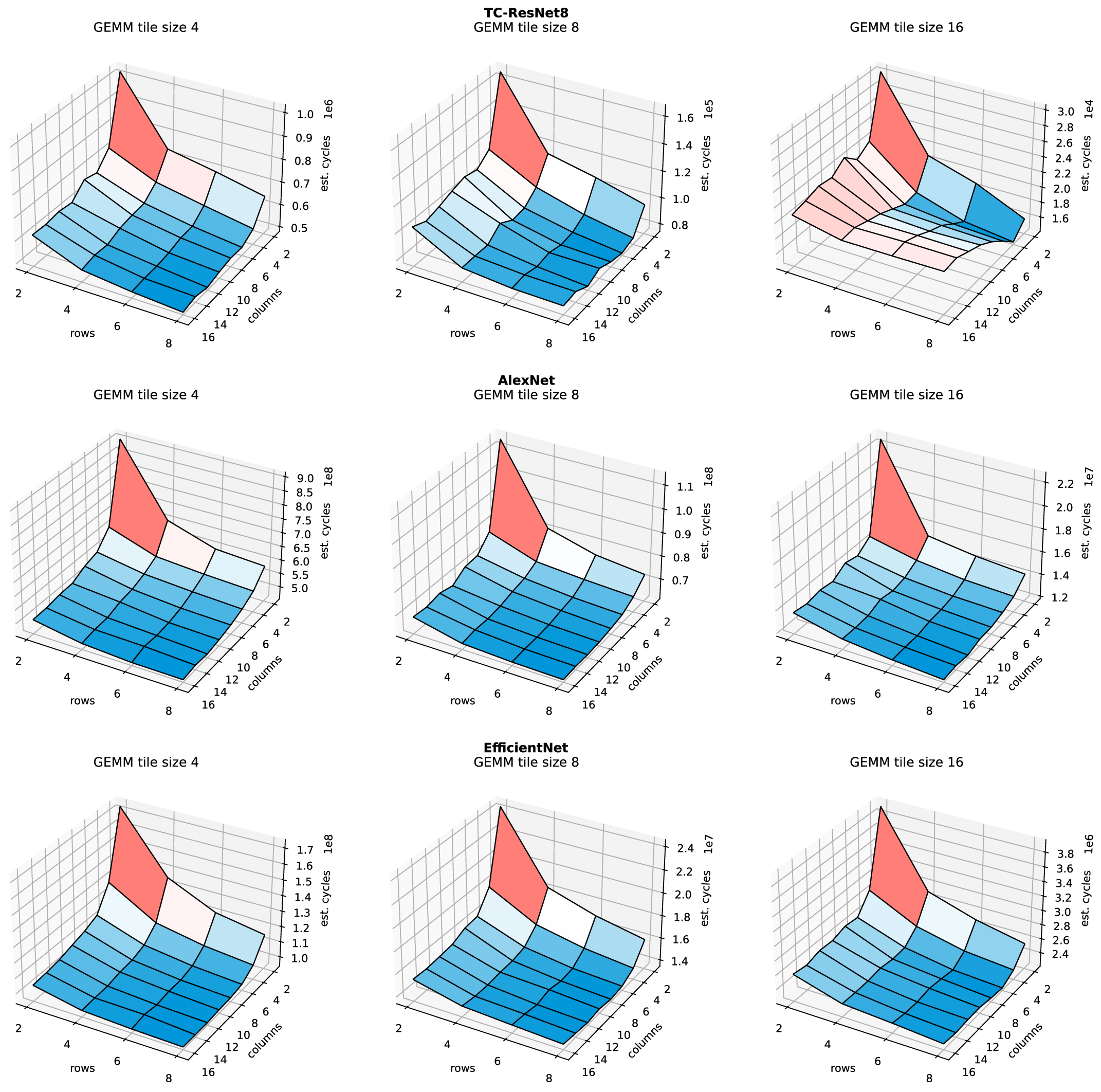}}
    \caption{Results of the design space exploration for mapping TC-ResNet8, AlexNet, and EfficientNet onto the different Plasticine-derived architecture instances.}
	\label{fig:plasticine_aidg_dse}
\end{figure}

Plasticine \cite{plasticine2017} is a parameterizable and reconfigurable architecture for parallel patterns developed at Stanford University. Plasticine comprises Pattern Compute Units (PCUs) and Pattern Memory Units (PMUs) arranged in a checkerboard pattern. PCUs and PMUs communicate via a switch box (S) interconnect, which can transfer data between units on a multi-word level. Fig.~\ref{fig:plasticine_block_diagram} shows a block diagram of the Plasticine-derived architecture. Inside each PCU resides a SIMD pipeline which can be configured to process parallel patterns such as Map, Fold, etc. PMUs have a scratchpad and a reconfigurable datapath that supports different access patterns. We opted to model a parameterizable Plasticine-derived architecture in ACADL on the matrix operation level. Each PCU is represented by an ExecuteStage, two RegisterFiles for input and output data, and a FunctionalUnit that supports tiled matrix-matrix multiplications (GEMM) and additions with fused activation and pooling operations. The PMUs are modeled using an ACADL Memory, an ExecuteStage, and MemoryAccessUnit, which supports instructions for read and write access to the Memory. The switch boxes comprise an ExecuteStage, a small RegisterFile, and a MemoryAccessUnit for moving data between PMUs, PCUs, and other switch boxes.

Using an im2col transformation for convolutional layer we can execute those layers as tiled GEMM and map them onto the PCUs of the Plasticine-derived architecture, while fully-connected layers can be executed directly without a transformation. We implemented a DNN mapper that maximizes the amount of parallel GEMM and matrix additions using our parameterizable ACADL model. Employing our mapper, we mapped the DNNs TC-ResNet8, AlexNet, and EfficientNet. We conducted an extensive design space exploration for all three DNNs where the rows, columns, and PCU GEMM tile size are varied to find the architectural parameters that result in the lowest end-to-end latency of a given DNN. 

Fig.~\ref{fig:plasticine_aidg_dse} presents the results of the design space exploration. Each row of plots presents the results for a DNN mapped onto the Plasticine-derived architecture. Going from left to right, the PCU GEMM tile size increases. The rows and columns of the Plasticine-derived architecture are varied within each plot. The z-axis of each plot presents the estimated cycles for a whole DNN for each parameter combination. One can see that a larger size and PCU GEMM tile size generally leads to a lower end-to-end latency -- except for TC-ResNet8 mapped onto the Plasticine-derived architecture with a PCU GEMM tile size of 16, where fewer rows and columns lead to the best latency. This is because the TC-ResNet8 layers are small and often fit into only a few tiled GEMM operations, such that the communication overhead to move the data into the PCUs takes longer than the tiled GEMM computation itself.

The mean runtime for mapping a DNN layer and the end-to-end latency estimation using our ACADL model and AIDG evaluation is 263.5\si{\second} with a standard deviation of 1379.6\si{\second}. This allows us to perform extensive design space explorations and speed up optimization loops in a hardware-aware NAS \cite{hannah2022}. Especially in the early design phase of an accelerator architecture, it is important to explore different design alternatives regarding memory bandwidth, data flow, and density of processing units to save on engineering time and cost. A fast design space exploration can help exclude design alternatives that do not yield optimal performance results and constrain the search space to find a close-to-optimal solution before writing RTL code.

\section{Conclusion}
This paper presented a fast and automatic approach for the generation of accurate performance models for DNN accelerator architectures based on the Abstract Computer Architecture Description Language (ACADL). Together with DNN layer mappings we construct and evaluate an Architectural Instruction Dependency Graph (AIDG) that allows us to evaluate, in the best case, only 154 loop kernel iterations to estimate the performance for 4.19 billion instructions. We evaluated our approach using four different accelerator architectures including UltraTrail, Gemmini, parameterizable systolic array, and Plasticine-derived. Those architectures are modeled on different abstraction levels going from fine-grained scalar operations up to fused tensor operations. We achieve better MAPE results than the state-of-the-art analytical latency estimators refined roofline and Timeloop and regression-based latency estimators without the need for creating a large training dataset.

We are confident that a fast and parameterizable modeling and performance estimation of DNN accelerator architectures is crucial for selecting a tailored hardware for the intended edge AI workload.

\begin{acks}
This work has been funded by the German Federal Ministry of Education and Research (BMBF) under grant number 16ES0876 (GENIAL!).
\end{acks}

\bibliographystyle{ACM-Reference-Format}
\bibliography{paper}

\appendix
\section{Appendix}
\subsection{Parameter Determination of AIDG Evaluation Fallback Heuristic}\label{sec:parameter_determination} 
To determine the percentage of iterations that are evaluated in the case of an oscillating $\Delta t_\text{iteration}$, we mapped the DNNs TC-ResNet8, AlexNet, and EfficientNet onto the systolic array, introduced in section \ref{sec:acadl_modeling_examples}, of different sizes (2$\times$2, 4$\times$4, 6$\times$6, 8$\times$8, and 16$\times$16). For each mapping of each DNN layer, we conducted an AIDG evaluation with a fixed point computation with different percentages of all iterations $k$ for the fallback heuristic. As percentages of $k$, we set 0.1\%, 1\%, and 5\% and compared the estimated end-to-end latency for all layers mappings which did not satisfy the fixed point criterion in equation (\ref{eqn:fixed_point_criterion}) against the AIDG whole graph evaluation of those layers using the MAPE. Additionally, we measured the runtime of the AIDG evaluation of each DNN layer mapping for the different percentages of $k$. 
\begin{figure}[htbp]
	\centerline{\includegraphics[width=1.0\linewidth]{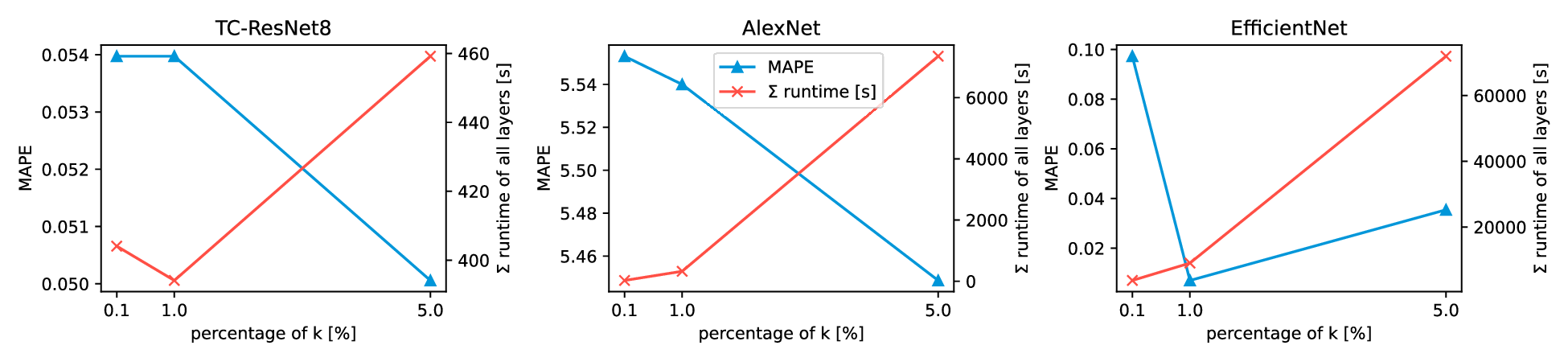}} 
	\caption{Comparison of different percentages of $k$ for the AIDG fixed point evaluation fallback heuristic of DNNs mapped on the systolic array of different sizes.}
	\label{fig:fallback_heuristic}
\end{figure}
Fig.~\ref{fig:fallback_heuristic} presents the MAPE and the runtime for each DNN mapped on systolic arrays of different sizes. For 0.1\% all DNN mappings show the highest MAPE. When increasing to 1\% the MAPE is either stable (TC-ResNet8) or decreases (AlexNet and EfficientNet). Further increasing to 5\% the MAPE drops significantly for TC-ResNet8 and AlexNet and increases slightly for EfficientNet. Comparing the estimation runtime of all DNN layers for different percentages of $k$ for the fallback heuristic, shows that a lower percentage of $k$ generally leads to a faster estimation runtime. This is because fewer iterations are evaluated in an AIDG until the fallback heuristic is applied. Therefore, the highest estimation runtime is measured for 5\%. For 1\% there is only a slight increase in the estimation runtime compared to 0.1\% for AlexNet and EfficientNet and even a slight drop for TC-ResNet8. Comparing the different percentages of $k$ shows that 1\% offers a good trade-off between estimation error and estimation runtime.

\begin{figure}[htbp]
	\centerline{\includegraphics[width=0.92\linewidth]{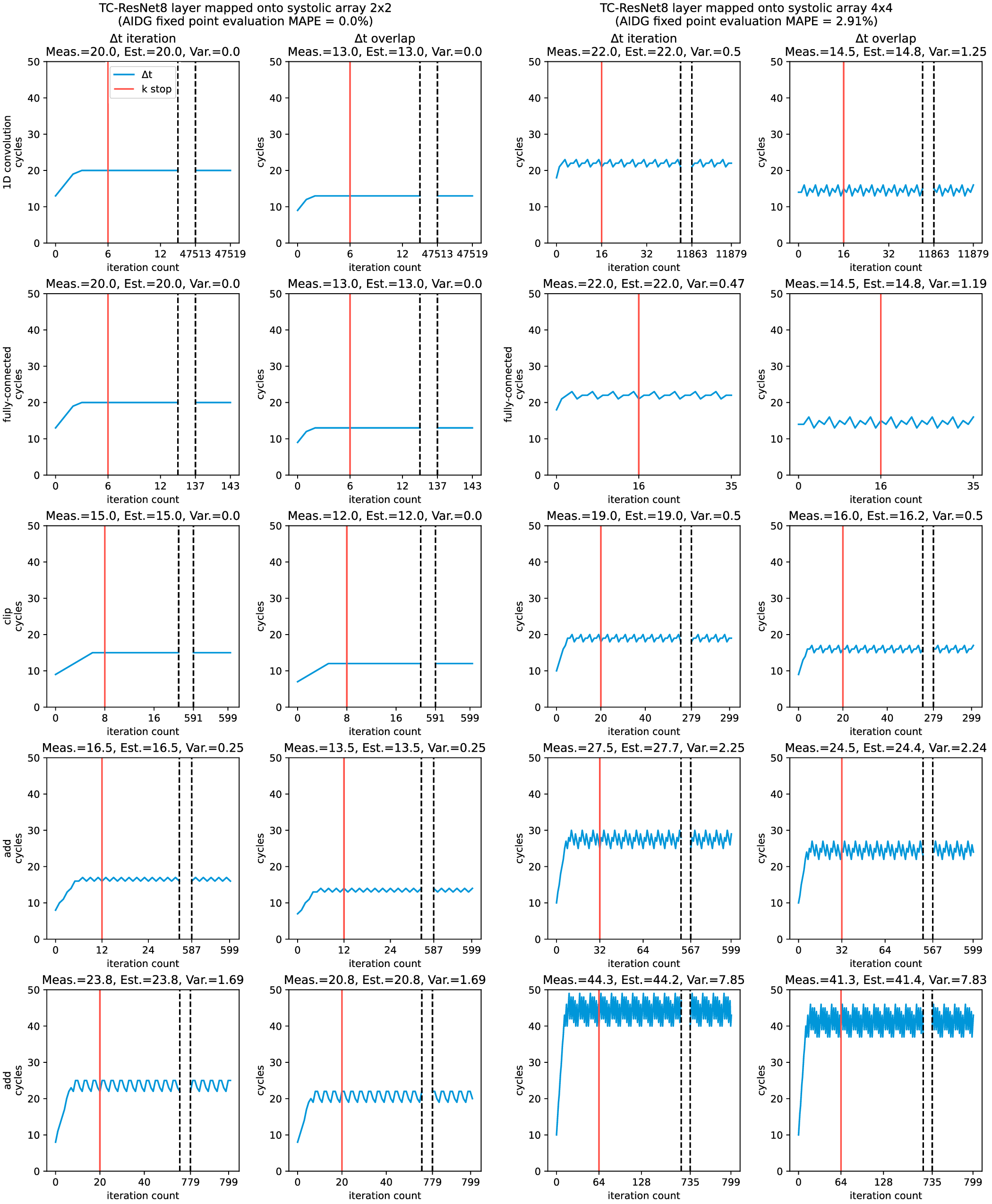}}
    \caption{Comparison of the variance $\Delta t_\text{iteration}$ and $\Delta t_\text{overlap}$ of different TC-ResNet8 layers mapped onto systolic array of size 2$\times$2 and 4$\times$4.}
	\label{fig:delta_t_iteration_overlap_variance_comparison}
\end{figure}

\subsection{Effects of Oscillating $\Delta t_\text{iteration}$ and $\Delta t_\text{overlap}$ on the Estimation Accuracy}\label{sec:premature_convergence}
To determine the effects of an oscillating $\Delta t_\text{iteration}$ and $\Delta t_\text{overlap}$ on the estimation accuracy of the AIDG fixed point evaluation we mapped the DNNs TC-ResNet8, AlexNet, and EfficientNet onto the systolic array of different sizes (2$\times$2, 4$\times$4, 6$\times$6, 8$\times$8, and 16$\times$16) using the same configurations as presented in Table~\ref{tab:systolic_array_results}. For each mapping, we recorded all $\Delta t_\text{iteration}$ and $\Delta t_\text{overlap}$ and did not stop the AIDG evaluation when equation (\ref{eqn:fixed_point_criterion}) was satisfied or 1\si{\percent} of all iterations $k$ have been evaluated.

Fig.~\ref{fig:delta_t_iteration_overlap_variance_comparison} presents the recorded $\Delta t_\text{iteration}$ and $\Delta t_\text{overlap}$ for different DNN layer types and configurations (1D convolution, clip, add, and fully-connected) of TC-ResNet8 mapped onto the systolic array of size 2$\times$2 and 4$\times$4. Each row of plots in Fig.~\ref{fig:delta_t_iteration_overlap_variance_comparison} corresponds to the same layer type and configuration of TC-ResNet8. The first two columns show the recorded $\Delta t_\text{iteration}$ and $\Delta t_\text{overlap}$ for the systolic array of size 2$\times$2, columns three and four show the recorded $\Delta t_\text{iteration}$ and $\Delta t_\text{overlap}$ for the systolic array of size 4$\times$4. The vertical red line marks the iteration $k_\text{stop}$ at which the AIDG fixed point evaluation would have stopped because either equation (\ref{eqn:fixed_point_criterion}) was satisfied or 1\si{\percent} of all iterations $k$ have been evaluated, according to the results presented in Table~\ref{tab:systolic_array_results}.

For the 1D convolutional and fully-connected layers in the first and second row of Fig.~\ref{fig:delta_t_iteration_overlap_variance_comparison} the number of iterations needed to process the layers on the systolic array of size 4$\times$4 is quartered compared to the number of iterations needed on the systolic array of size 2$\times$2. This is because 1D convolutional and fully-connected layers have high data reuse and the input channel dimensions of those layer configurations are divisible by 2 and 4 which allows for an optimal utilization of processing elements. The number of iterations of the clip layer presented in the third row of Fig.~\ref{fig:delta_t_iteration_overlap_variance_comparison} is halved when increasing the systolic array size from 2$\times$2 to 4$\times$4 because the input channel dimension is divisible by 2 and 4. Moreover, the clip layer is applied element-wise and does not have any data reuse. Therefore, only the first row of processing elements of the systolic array is utilized, which corresponds to a linear decrease in the number of iterations when increasing the systolic array size. The input channel dimensions of the add layers in the fourth and fifth row of Fig.~\ref{fig:delta_t_iteration_overlap_variance_comparison} are not divisible by 2 or 4. Therefore, only one processing element is utilized in the systolic array of size 2$\times$2 and 4$\times$4. The contribution of the add layers to the total execution time of the DNN is negligible, and therefore a non-optimal mapping is acceptable. Fig.~\ref{fig:delta_t_iteration_overlap_variance_comparison} also shows that non-optimal mappings, such as the mapping of the add layer, lead to a stronger oscillation of $\Delta t_\text{iteration}$ and $\Delta t_\text{overlap}$. Furthermore, this showcases that our proposed execution time estimation approach can also make accurate estimations for non-optimal mappings in contrast to other estimation methods which assume an optimal or constant utilization. 

$\Delta t_\text{iteration}$ and $\Delta t_\text{overlap}$ is increased for all layers in Fig.~\ref{fig:delta_t_iteration_overlap_variance_comparison} when going from a systolic array of size 2$\times$2 to 4$\times$4 because the data must vertically pass through double the number of processing elements to reach the store units at the bottom of the systolic array.

For each evaluated DNN layer $i$ mapped onto all systolic array sizes, we calculated the sample variance of $\Delta t_\text{iteration}$ and $\Delta t_\text{overlap}$ from $k_\text{stop}$ (AIDG fixed point evaluation stop) to $k$ (all iterations). Afterward, we calculated the mean sample variance over all layers for each DNN mapping using the following equations:
\begin{equation}
    \label{eq:var_iteration}
    \meanvardeltatiteration = \frac{1}{n}\sum_{i=0}^n\operatorname{Var}\left(\Delta {{t_\text{iteration}}_i}_{[k_\text{stop},k]}\right)
\end{equation}
and
\begin{equation}
    \label{eq:var_overlap}
    \meanvardeltatoverlap = \frac{1}{n}\sum_{i=0}^n\operatorname{Var}\left(\Delta {{t_\text{overlap}}_i}_{[k_\text{stop},k]}\right).
\end{equation}
The mean sample variance of a mapping characterizes how much $\Delta t_\text{iteration}$ and $\Delta t_\text{overlap}$ for a whole DNN oscillate after the AIDG fixed point evaluation stopped. Table~\ref{tab:systolic_array_mape_vs_variance} shows the mean sample variance of $\Delta t_\text{iteration}$ and $\Delta t_\text{overlap}$ for each DNN systolic array mapping. Additionally, Table~\ref{tab:systolic_array_mape_vs_variance} shows the percentage of DNN layers of a mapping for which equation (\ref{eqn:fixed_point_criterion}) was not satisfied and the fallback heuristic was used.

The percentage of layers for which the fallback heuristic was applied increases with the systolic array size because in general, the larger a systolic array is, the fewer iterations are needed to process a layer (see Fig.~\ref{fig:delta_t_iteration_overlap_variance_comparison}). Having less iterations in a layer mapping will trigger the fallback heuristic earlier than the fixed point criterion being satisfied. Secondly, Table~\ref{tab:systolic_array_mape_vs_variance} shows that the mean sample variance of $\Delta t_\text{iteration}$ and $\Delta t_\text{overlap}$ between $k_\text{stop}$ and $k$ increases with the size of the systolic array because there are more structural and data dependencies, leading to more stalls due to a more complex hardware structure involving more hardware modules.  

\begin{table}[htbp]
    \caption{Comparison of MAPE, mean sample variance of $\Delta t_\text{iteration}$ and $\Delta t_\text{overlap}$, and the percentage of DNN layers for which the fallback heuristic was used for different DNNs mapped onto systolic array instances of different sizes.}
	\label{tab:systolic_array_mape_vs_variance}
	\scriptsize{
	\begin{center}
	\begin{tabular}{clccccc}
	\toprule
        Size & DNN & MAPE & $\meanvardeltatiteration$ & $\meanvardeltatoverlap$ & layers estimated with fallback heuristic\\
	\midrule
        \multirow{3}*{2$\times$2}   & TC-ResNet8    & \num{0.0}\si{\percent}    & \num{0.3}     & \num{0.15} & \num{0.0}\si{\percent} \\
                                    & AlexNet       & \num{0.0}\si{\percent}    & \num{0.23}    & \num{0.11} & \num{0.0}\si{\percent} \\
                                    & EfficientNet  & \num{0.0}\si{\percent}    & \num{5.44}    & \num{3.67} & \num{6.35}\si{\percent} \\
	\addlinespace
        \multirow{3}*{4$\times$4}   & TC-ResNet8    & \num{2.91}\si{\percent}   & \num{1.18}   & \num{1.53} & \num{0.0}\si{\percent} \\
                                    & AlexNet       & \num{2.59}\si{\percent}   & \num{0.35}   & \num{0.57} & \num{0.0}\si{\percent} \\
                                    & EfficientNet  & \num{1.73}\si{\percent}   & \num{6.74}   & \num{8.39} & \num{15.3}\si{\percent} \\
    \addlinespace
        \multirow{3}*{6$\times$6}   & TC-ResNet8    & \num{0.01}\si{\percent}   & \num{1.34}   & \num{1.38} & \num{29.17}\si{\percent} \\
                                    & AlexNet       & \num{0.01}\si{\percent}   & \num{0.58}   & \num{0.6}  & \num{22.22}\si{\percent} \\
                                    & EfficientNet  & \num{0.36}\si{\percent}   & \num{14.36}  & \num{13.69} & \num{27.84}\si{\percent} \\
    \addlinespace
        \multirow{3}*{8$\times$8}   & TC-ResNet8    & \num{2.65}\si{\percent}   & \num{2.69}   & \num{2.36} & \num{26.09}\si{\percent} \\
                                    & AlexNet       & \num{7.1}\si{\percent}    & \num{0.82}   & \num{0.58} & \num{0.0}\si{\percent} \\
                                    & EfficientNet  & \num{1.17}\si{\percent}   & \num{10.04}  & \num{8.42} & \num{29.7}\si{\percent} \\
    \addlinespace
       \multirow{3}*{16$\times$16}  & TC-ResNet8    & \num{0.31}\si{\percent}    & \num{0.8}   & \num{0.75} & \num{72.73}\si{\percent} \\
                                    & AlexNet       & \num{5.15}\si{\percent}   & \num{0.97}  & \num{0.84} & \num{55.56}\si{\percent} \\
                                    & EfficientNet  & \num{1.62}\si{\percent}   & \num{12.97}  & \num{11.69} & \num{65.85}\si{\percent} \\
	\bottomrule 
	\end{tabular}
	\end{center}
    }
\end{table}

Table~\ref{tab:systolic_array_mape_correlation} shows the pearson correlation coefficient $\rho$ of the MAPE compared to the mean sample variance of $\Delta t_\text{iteration}$ and $\Delta t_\text{overlap}$, and the percentage of layers which were estimated using the fallback heuristic. For TC-ResNet8 and AlexNet there is a strong correlation between the mean sample variance of $\Delta t_\text{iteration}$ and $\Delta t_\text{overlap}$ and the MAPE while the same correlation is not as strong for EfficientNet. However, the EfficientNet layers contain a low number of iterations which leads to a higher percentage of layer for which the fallback heuristic was used as shown in Table~\ref{tab:systolic_array_mape_vs_variance}. Furthermore, the correlation between the MAPE and the percentage of layers for which the fallback heuristic was used for EfficientNet is high with a value of 0.52.

\begin{table}[htbp]
    \caption{Pearson correlation coefficient $\rho$ for the MAPE, mean sample variance of $\Delta t_\text{iteration}$ and $\Delta t_\text{overlap}$, and the percentage of layers for which the fallback heuristic was used.}
	\label{tab:systolic_array_mape_correlation}
	\begin{minipage}{\columnwidth}
	\footnotesize{ 
	\begin{center}
	\begin{tabular}{l ccc}
	\toprule
        DNN & $\rho\left(\text{MAPE}, \meanvardeltatiteration\right)$ & $\rho\left(\text{MAPE}, \meanvardeltatoverlap\right)$ & $\rho(\text{MAPE}, \text{fallback heuristic})$\\
	\midrule
        TC-ResNet8  & \num{0.65}    & \num{0.75}    & \num{-0.28} \\ 
        AlexNet     & \num{0.76}    & \num{0.57}    & \num{0.18} \\ 
        EfficientNet & \num{0.1}    & \num{0.3}     & \num{0.52}\\ 
	\bottomrule 
	\end{tabular}
	\end{center}
	}
	\end{minipage}
\end{table}

\begin{figure}[htbp]
	\centerline{\includegraphics[width=1.0\linewidth]{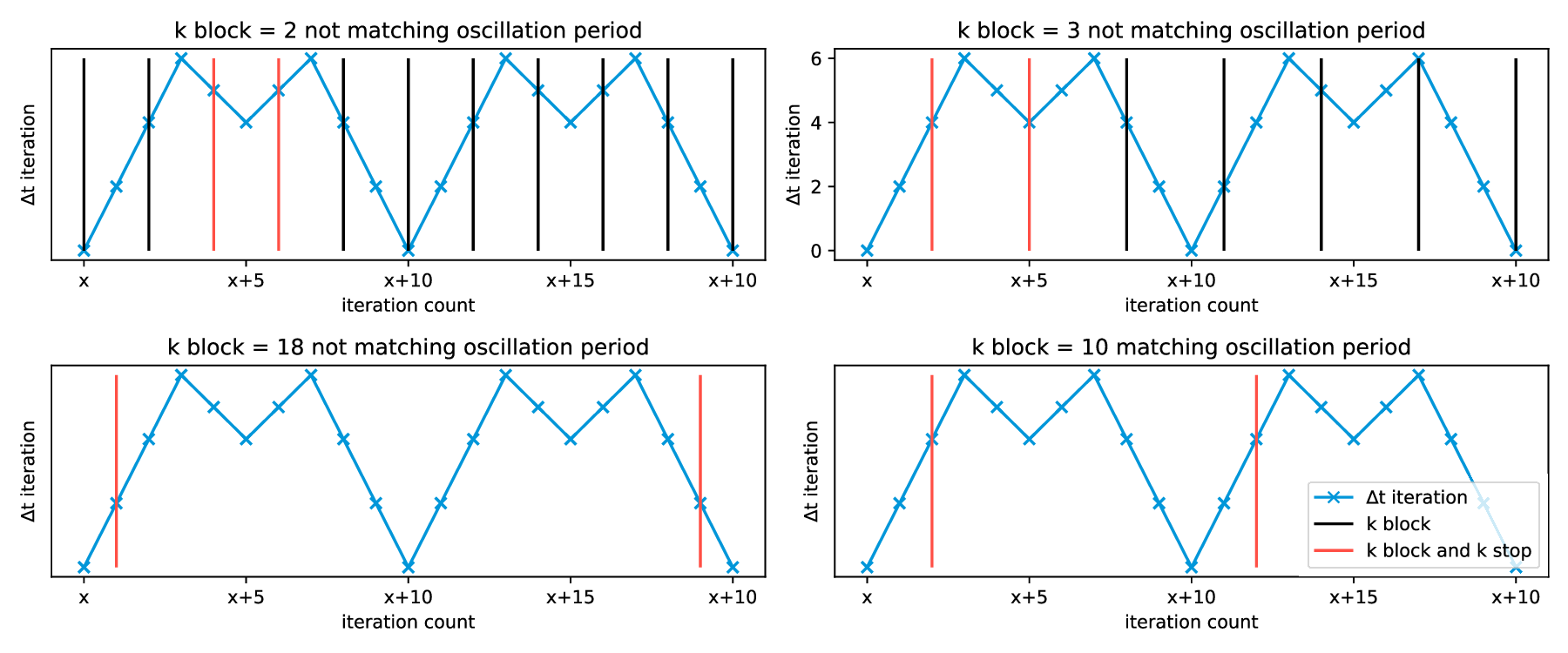}}
    \caption{Illustration of different $k_\text{block}$ lengths applied to the same oscillating $\Delta t_\text{iteration}$. The black lines show the end of $k_\text{block}$ while the red lines show the ends of two consecutive blocks of iterations of length $k_\text{block}$ whose $\Delta t_\text{iteration}$ satisfy equation (\ref{eqn:fixed_point_criterion}).}
	\label{fig:delta_t_k_block_plot}
\end{figure}

The mean sample variance of $\Delta t_\text{iteration}$ and $\Delta t_\text{overlap}$ and the percentage of layers which were estimated using the fallback heuristic explains why the MAPE for estimations done on the systolic array instances of size 4$\times$4 and 8$\times$8 is higher compared to the instances of size 2$\times$2, 6$\times$6, and 16$\times$16. Firstly, our proposed methodology for detecting a fix-point using equation (\ref{eqn:fixed_point_criterion}) only incorporates two consecutive blocks of iterations. The length of a block of iterations $k_\text{block}$ does not necessarily match the period of the oscillating $\Delta t_\text{iteration}$ and $\Delta t_\text{overlap}$. Fig.~\ref{fig:delta_t_k_block_plot} illustrates four different $k_\text{block}$ lengths applied to an oscillating $\Delta t_\text{iteration}$ with a period of 10 iterations. Even though the oscillation period is 10, equation (\ref{eqn:fixed_point_criterion}) is also satisfied with a $k_\text{block}$ length of e.g. 2, 3, and 18 which leads to a slight over- or underestimation of $\Delta t_\text{iteration}$. This estimation error is correlated to the mean sample variance of $\Delta t_\text{iteration}$ and $\Delta t_\text{overlap}$ because a higher variance can lead to a higher estimation error as the estimated $\Delta t_\text{iteration}$ and $\Delta t_\text{overlap}$ from the fixed-point evaluation assume a constant $\Delta t_\text{iteration}$ and $\Delta t_\text{overlap}$.

Secondly, the fallback heuristic uses the mean $\Delta t_\text{iteration}$ taken from a sample of evaluated iterations, which can also contribute to an estimation error. The assumption that $\Delta t_\text{iteration}$ and $\Delta t_\text{overlap}$ are constant after $k_\text{stop}$ allows for a fast performance estimation using the fixed-point evaluation or the fallback heuristic, which also provides high accuracy compared to other performance estimation approaches.

\end{document}